# Comparing GNSS Derived Sea Ice Drift in the Arctic and Antarctic using Rotary Spectra and Principal Component Analysis

**James H. Hepworth**[1*] · **Amit Kumar Mishra**[2,3]

[1]Department of Mechanical Engineering, University of Cape Town, Cape Town, South Africa

[2]University West, Sweden

[3]Faculty of Space Technology, AGH University of Krakow, Poland

[*]Corresponding author: james.hepworth@uct.ac.za

**Abstract** Sea ice drift underpins air–sea ice coupling and model evaluation, but the Southern Ocean remains chronically undersampled relative to the Arctic. This work presents a cross-polar comparison of GNSS-tracked ice-drift time series using rotary spectral analysis, principal component analysis, inter-buoy coherence, complementary shape- and amplitude-sensitive spectral differences. Arctic data are aggregated into Beaufort Gyre and Transpolar Drift seasonal composites (2017–2024); Antarctic data comprise eight heterogeneous campaigns (2000–2022) with small buoy counts and short durations. After common quality control, the clearest cross-polar similarity is spectral organisation: both regions show strong low-frequency variance and a hemisphere-appropriate rotary enhancement near the Coriolis frequency, interpreted cautiously as a combined inertial–semidiurnal response. Coherence is highest at synoptic scales and declines toward higher frequencies. Southern Ocean campaigns occupy a higher-energy envelope, so similar band structure does not imply similar drift amplitude. Principal components capture much array-scale motion, but high cumulative variance does not substitute for dense sampling of smaller-scale processes. The results provide transferable frequency-resolved benchmarks (dominant variance below 0.5 cpd and in the near-inertial band, an expanded Arctic reference, and explicit separation of spectral shape from amplitude) to inform future observing-system design without assuming identical dynamical regimes.

---

## 1 Introduction

The motion of sea ice across the Earth's polar oceans influences a wide range of systems. These range from the global energy balance at the air–sea ice interface to operational planning for marine vessels in polar waters. Ice motion is also a fundamental feature in numerical ocean models that help us understand and evaluate oceanic processes and the effects they have on the broader world [1]. Traditionally, sea ice drift has been described as a function of low-frequency and high-frequency components [2], [3]: low-frequency drivers force slow changes in ice motion at the seasonal scale or longer [3], [4], [5], while high-frequency drivers then force variation on this mean motion, with complex, multi-fractal behaviour driven by many transient effectors [3], [6], [7]. These drivers include ocean currents, inertial oscillations, tides, and winds, which operate over a variety of time-scales. Ocean currents force slow changes in the motion, while tides, inertial oscillations and winds account for daily and hourly variations, respectively.

Further, the interactions between ice floes themselves cause higher frequency variations in the motion as a result of inter-floe collisions.

Drift itself is monitored using a variety of technologies and approaches. It is directly assessed by ice-tethered platforms that track ice-floe locations using Global Navigation Satellite Systems (GNSS). Satellite remote sensing in the infra-red [8] and microwave [9] bands, as well as ship-borne and synthetic aperture radar (SAR) systems [1], [10], have also been used.

While the motion of ice-drift occurs in both the Arctic and the Southern Ocean (SO), it has been much better characterised in the Arctic than the SO due to the significantly greater quantity of data available from the Arctic than the SO. The Arctic Ocean is a semi-enclosed basin dominated by the Beaufort Gyre (BG) and Transpolar Drift (TD) Stream currents. The continental land masses of North America, Europe and Asia bound the Arctic on most sides with small exit regions in the Fram Strait between Greenland and Svalbard, and the Bering Strait between Russia and Alaska. By contrast the SO is a far more exposed ocean surrounding the Antarctic continent. The Antarctic Circumpolar Current circulates anticlockwise around the continent, driven by the strong winds blowing over the Southern Ocean with effectively unbounded fetch. This current is the dominant driver of sea ice drift in the SO, with the ice being advected eastward by the current. With the increased remoteness of the SO from the inhabited continents, and the resultant difficulty of access and the prohibitively high cost to do so [11], [12], the SO has been chronically under-sampled in terms of in-situ, spatially dense, long time series observational data [13]. Furthermore, deploying in-situ sensors and ensuring their survivability is difficult, with wave action near the sea ice edge having been shown to reach amplitudes of between four and five metres [14].

These challenges make it difficult to validate scientific models, remote sensing packages, and reanalysis products for the SO. Therefore, there are large variations and uncertainties associated with products' outputs, including coupled climate and weather forecasting models, flux and wave climatology, and atmospheric reanalysis products [15], [16], [17]. Thus, there is a clear need for increased in-situ observations from the SO in the future. However, with a comparatively short observational record from the SO the choice of where to deploy these platforms and how to sample them such that maximum value from their measurements can be derived is non-trivial.

While deployment locations for ocean buoys may be chosen according to the requirements set by a Global or Regional Basic Observing Network (GBON or RBON) standard in place for a set of measurement parameters, for research experiments, such as those prevalent in the SO, the locations of new ocean buoys are usually chosen through the use of an Observing System Simulation Experiment (OSSE) [18], [19]. These are simulations used to investigate the potential impact of a proposed observing system on the performance of a model representing the environmental process of interest. Usually, these are numerical models which are provided with synthetic data from the proposed observing system which has been generated using reanalysis and remote sensing products. Numerical models providing sea ice parameters used in these OSSEs include the Regional Ocean Modelling System (ROMS) and the Nucleus for European Modelling of the Ocean (NEMO) models. However, neither of these models explicitly track the uncertainty or error of their parameters and, given the wide uncertainties associated with reanalysis and remote sensing products, OSSEs remain difficult to validate in the development of deployment strategies for new sensors in the SO. The synthetic data derived from reanalysis products are uncertain, and the numerical models do not adequately track error, so any assessed change in model performance has low accuracy. In contrast to the SO, due to the rich observational record (e.g., in-situ, remote sensing, and validated climate products), reanalysis and remote sensing products in the Arctic are well validated and thus OSSEs are more reliable tools for the investigation of new observing systems.

The purpose of this study is to investigate similarities and differences in the in-situ drift record from both the Arctic and the SO with a view to informing the design and sampling strategies for future observing systems in the SO. Other studies in the regions have investigated sea ice drift as it relates to descriptors of motion such as dispersion, drift speed, turning angles and deformation rates [20], [21], [3], [22]. Further, drift has been studied as it relates to the underlying physical drivers of motion such wind, waves, tidal forcing, and Coriolis, using a variety of methods including dispersion, strain rate, and wavelet analysis [23], [5], [24], [25]. In this study we take an approach that is deliberately agnostic at the outset to both the underlying physics of motion and the usual descriptors used to characterise sea ice drift. Instead, we concentrate on the time-series positions of deployed buoys in the Arctic and in the SO in order to assess the similarity of drift between the two oceans from a signal-processing perspective to inform engineering design and optimisation of future observing systems.

To achieve this, the following methods are employed: First, the Arctic and the SO are each evaluated using (1) rotary spectral analysis to decompose the drift into its frequency domain representation so as to assess the time-scale structure of drift and the energy distribution across the synoptic, inertial, and higher-frequency scales. This is followed by (2), principal component analysis (PCA) to identify the proportion of total variance captured by the principal modes of drift and to assess modal compactness and modal frequency content. Given the quantity and richness of the Arctic data, the Arctic analysis is done on a seasonal basis spanning the years 2017–2024 to present a climatological view of the Arctic sea ice drift over the previous eight years of available data. In contrast, the Antarctic analysis is done on a campaign basis spanning the years 2000-2022 and is presented in terms of individual campaigns. Finally, (3) an inter-ocean comparison is performed to assess the similarity of drift in the SO campaigns compared to the Arctic seasonal reference.

The contribution of the study is therefore a rotary-spectral, dimensionality and coherence-based cross-polar comparison to identify transferable design parameters for future observing systems: an expanded Arctic seasonal reference, a set of frequency-resolved shape and amplitude benchmarks against which sparse Antarctic campaigns can be assessed, and an explicit separation of spectral-shape similarity from absolute energy-amplitude similarity that guides sampling-rate and buoy-spacing choices.

This analysis is presented in the following sections: First, the data sources, their grouping, and preprocessing are described in Section 2. Subsequently the methodology employed in this analysis is presented in detail in Section 3. This is followed first, by the presentation of the Arctic analysis results in Section 4 and then by the SO campaigns analysis results in Section 5. These are then compared in Section 6 to highlight the similarity and differences between drift in the two oceans before the contributions of this study are discussed in Section 7. The paper is then concluded in Section 8.

## 2 Data Description

### 2.1 Data sources and grouping

This investigation employs sea ice drift data derived from the GNSS locations of ice-tethered ocean buoys in the Arctic and the Antarctic. The measurement platforms in this study are various buoy hardware systems deployed in the Arctic and the Antarctic over the last 20 years. These buoys are typically designed to gather met-ocean parameters such as atmospheric temperature and pressure, sea-surface temperature and GNSS position. Several are also equipped with Inertial Measurement Units (IMUs) to measure wave action in the ice fields. Architecturally, they are similar to those described in [26]. Reported data do not typically include the dilution of precision metric of the GNSS data in order for an error analysis to be performed specific to each individual experiment. However, other work has shown that standalone GNSS positioning at latitudes greater than 60° may exhibit error of a magnitude up to 10 m [27]. In the data sets used, positions are reported frequently to a precision of $0.1 \times 10^{-}3^\circ$ ( 11 m) in the latitude and longitude directions. Accordingly, data from sets reporting precision greater than this threshold are rounded to the fourth decimal place of the degree in analysis for consistency.

In the analysis that follows, rich quantities of Arctic data were available spanning many years and with high spatial and temporal resolution. In this work, therefore, the Arctic data are used to identify the quasi-climatological structure of the Arctic sea ice drift according to the methodology that follows in the next section. Data are grouped by ocean current dominated regimes based on the buoy deployment locations; either the BG or TD. In contrast, the Antarctic data for sea ice drift are extremely sparse and heterogeneous in terms of deployment geometry, experiment length and in the number of buoys deployed.

**Table 1** data set metadata summarising the Arctic and Antarctic data used in this study. Native sampling and analysis grid describe the uniform grids used after preprocessing; ensemble statistics (buoys, duration, mean position) are from the processed records used here.

| Experiment | Region | Datasets | Year(s) | Mean buoy count | Mean duration (days) | Mean latitude (°) | Mean longitude (°) | Native sampling | Analysis grid | Source |
|---|---|---|---|---|---|---|---|---|---|---|
| BG (all grouped datasets) | BG | 32 | 2017–2024 | 58.06 | 66.92 | 76.62 | −148.67 | 30 min–6 h | 1 h | [28] [29] |
| TD (all grouped datasets) | TD | 32 | 2017–2024 | 133.28 | 67.52 | 78.91 | 3.91 | 30 min–6 h | 1 h | [28] [29] |
| SHARC 2022 | Antarctic | 1 | 2022 | 3.00 | 1.74 | −58.78 | −0.48 | 15 min | 15 min | *Internally collected data* |
| SCALE SPRING 2019 | Antarctic | 1 | 2019 | 3.00 | 50.88 | −58.34 | 8.92 | 1 h | 1 h | [30] |
| SCALE WINTER 2019 | Antarctic | 1 | 2019 | 3.00 | 27.92 | −56.34 | 4.43 | 1 h | 1 h | [30] |
| PIPERS 2017 | Antarctic | 1 | 2017 | 6.00 | 8.84 | −66.95 | 181.29 | 15 min | 15 min | [31] [32] |
| WIIOS 2017 | Antarctic | 1 | 2017 | 2.00 | 8.76 | −62.47 | 30.74 | 15 min | 15 min | [33] [34] |
| SIPEX II BUOYS 2012 | Antarctic | 1 | 2012 | 8.00 | 17.50 | −64.62 | 116.50 | 15–45 min | 15 min | [35] [26] |
| SIPEX II WAVES 2012 | Antarctic | 1 | 2012 | 4.00 | 8.12 | −61.85 | 122.47 | 3 h | 3 h | — |
| STiMPI 2000 | Antarctic | 1 | 2000 | 6.00 | 17.94 | −68.45 | −33.35 | 20 min | 20 min | [21] |

Accordingly, they are analysed and presented as individual campaigns and are then compared with the Arctic climatological structure. These campaigns are listed alongside the Arctic data in Table 1 above and are summarised in further detail in Table 4 of Section 5. Here it can be seen that BG averages about 58 buoys per grouped data set and TD about 133, both over roughly 67 days, whereas Antarctic campaigns contain only 2–8 buoys and, on the common overlapping analysis window used here, span about 1.7–50.9 days.

## 2.2 Data preprocessing

As shown in Table 1, the GNSS positions from the various buoys used in this study were sampled at different frequencies, with nominal sampling periods ranging from 15 minutes to 6 hours. Most raw data sets contained gaps where GNSS locations could not be determined for several consecutive sampling periods. Since both the rotary spectral analysis and time-domain PCA used in the sections that follow require constant sampling rates, preprocessing of the data was necessary. This preprocessing ensured that position measurements were aligned to comparable time instances and that each data set maintained entries at uniform sample rates corresponding to the modal sampling frequency of each experimental data set. The raw data sets were thus preprocessed using linear interpolation to fill missing data points for time gaps not exceeding 24 hours in duration.

The BG and TD datasets draw records from five source programmes: The International Arctic Buoy Programme (IABP) [36], Ice-Tethered Profilers (ITP) [28], [29], the Multidisciplinary drifting Observatory for the Study of Arctic Climate (MOSAiC) [37], PS131 NOMAD [38], and the Seasonal Ice Zone Experiment (SIDEx) [39],[40]. Natively, these data products are presented with different temporal resolutions, file types and metadata. To construct the seasonal, region based ensemble datasets used in this work, the source data were cleaned by converting data to standardised timestamps, longitude format using the −180° to +180° range with cumulative phase unwrapping where tracks cross the 180° meridian, removal of missing or flagged positions, and step-based quality control: Buoys were excluded from the record entirely if any single inter-point step exceeded a distance of 20 km (as an indicator of gross position error rather than physical drift). Temporal resolution was not uniform between sources: IABP and ITP generally provide hourly or better sampling, whereas MOSAiC and SIDEx include mixed intervals that needed reconciling before inclusion in the ensemble datasets. To facilitate the seasonal Arctic composites, Arctic records were resampled to a common, one-hour sampling rate. Following this, the records were then assigned to meteorological seasons to create a set of regime-season date for use in the analysis. Winter is defined here as from December of a previous year to February of the year in question, Spring as March–May, Summer as June–August, and Autumn as September–November of the year in question).

Antarctic campaigns were instead analysed on their campaign-specific uniform grids, which range from 15 minutes to 3 hours after preprocessing (Table 1). Consequently, high-frequency interpretation is constrained by the native sampling of each campaign: hourly records have a 12 cycles per day (cpd) Nyquist frequency, 15 minute records a 48 cpd Nyquist frequency, 20 minute records a 36 cpd Nyquist frequency, and the 3 hour SIPEX II Waves record a 4 cpd Nyquist frequency. Therefore, the results that follow in this paper emphasise variability that is resolved on the relevant analysis grid, and cross-regional comparison in Section 6 are evaluated only over frequency bins shared by the spectra being compared.

## 3 Methodology

In this work, sea ice drift in the Arctic and Antarctic is analysed and compared through two primary approaches: rotary spectral analysis and Principal Component Analysis (PCA). Rotary spectral analysis is used to decompose drift into its frequency-domain representation to assess the time-scale structure of drift and the energy distribution across these scales. Rotary analysis distinguishes between clockwise and counter-clockwise components of motion present in the drift velocity vector. PCA is then used to identify the principal modes of drift variability and to assess modal compactness and modal frequency content.

Throughout, horizontal drift is treated as a two-component vector process so that rotary spectra, coherence, and PCA are applied to physically comparable quantities rather than to scalar displacement alone.

### 3.1 Regime-level analysis

The regime-level analysis comprises the derivation of buoy velocity components, rotary spectra, uncertainty bands based on ensemble spread, and principal-component summaries for each Arctic regime-season composite and each Antarctic campaign. These steps define the within-regime and within-campaign behaviour that is later compared between polar settings.

#### 3.1.1 Displacement, velocity, and time alignment

For any buoy, consecutive latitude-longitude pairs are first converted from degrees to radian-valued coordinates $(\varphi_{i-1}, \lambda_{i-1})$ and $(\varphi_i, \lambda_i)$, from which the great-circle step length $d_i$ is computed using the Haversine formula with Earth radius $R = 6\,371\,\mathrm{km}$,

$$d_i = 2R \arcsin\left(\sqrt{\sin^2\left(\frac{\Delta\varphi_i}{2}\right) + \cos(\varphi_{i-1})\cos(\varphi_i)\sin^2\left(\frac{\Delta\lambda_i}{2}\right)}\right) \tag{1}$$

where $\Delta\varphi_i = \varphi_i - \varphi_{i-1}$, $\Delta\lambda_i = \lambda_i - \lambda_{i-1}$, and the first displacement is set to zero. Eastward and northward velocity components, $u_i$ and $v_i$, respectively, are then obtained from the original degree-valued coordinate differences using a local tangent-plane approximation of 111 km per degree of latitude with $\Delta t_i$ the elapsed time in seconds:

$$\bar{\varphi}_i = \frac{\varphi_i + \varphi_{i-1}}{2}, \quad \Delta x_i = (\lambda_i^\circ - \lambda_{i-1}^\circ)(111\,000\,\mathrm{m\,deg^{-1}})\cos\bar{\varphi}_i, \quad \Delta y_i = (\varphi_i^\circ - \varphi_{i-1}^\circ)(111\,000\,\mathrm{m\,deg^{-1}}) \tag{2}$$

$$u_i = \frac{\Delta x_i}{\Delta t_i}, \quad v_i = \frac{\Delta y_i}{\Delta t_i} \tag{3}$$

Furthermore, all buoys employed in the analysis are restricted to the overlapping interval $[T_{\mathrm{start}}, T_{\mathrm{end}}]$ for each regime-season (in the Arctic) or campaign (in the SO) with $T_{\mathrm{start}} = \max_i \text{start-time}_i$ and $T_{\mathrm{end}} =$ $\min_i$ end-time$_i$. Whilst this is not strictly necessary for the rotary spectral analysis, it is necessary for the PCA analysis because PCA is applied to a common-time matrix of buoy velocity components.

### 3.1.2 Rotary spectral analysis

Following the approach introduced in [41], [42] (and others), rotary power spectral density (PSD) estimates are computed from the complex velocity series $z[n] = u[n] + \mathrm{j}\, v[n]$, where $n = 1, ..., N$ is the time index of the number of samples in the time series, $N$. Prior to PSD estimation, a Hamming window is applied to each real component to suppress spectral leakage from finite records while retaining enough frequency resolution to identify broad low-frequency slopes and near-Coriolis enhancements.[1]:

$$w[n] = 0.54 - 0.46 \cos\left(\frac{2\pi n}{N-1}\right), \quad n = 0, ..., N-1. \tag{4}$$

Accordingly, the rotary spectrum coefficients $Z[k]$ are computed as:

$$Z[k] = \sum_{n=0}^{N-1} z[n]\, w[n]\, e^{-\mathrm{j}2\pi k \frac{n}{N}} = U[k] + \mathrm{j}\, V[k], \tag{5}$$

where $U[k]$ and $V[k]$ are the windowed DFTs of $u[n]$ and $v[n]$ individually. The windowed rotary periodogram in Hertz is estimated as

$$S_{\mathrm{Hz}}[k] = \frac{|Z[k]|^2}{N f_s}, \tag{6}$$

where $f_s = \frac{1}{\Delta t}$ is the sampling rate in Hertz and $\Delta t$ is the uniform time step after preprocessing. This common normalisation is applied to all spectra. Because the Hamming-window power is not separately used to rescale the periodogram, absolute integrated values should be interpreted as consistently windowed PSD estimates rather than as independently calibrated kinetic-energy budgets. Frequencies are expressed in cpd, with $f_{\mathrm{cpd}} = f_{\mathrm{Hz}} \times 86\,400$. For a spectral density with respect to $f_{\mathrm{cpd}}$, values used in figures and exported spectra are

$$S_{\mathrm{cpd}} = \frac{S_{\mathrm{Hz}}}{86\,400}, \tag{7}$$

with units $(\mathrm{m/s})^2/\mathrm{cpd}$.

Calculation of the rotary spectra allows the identification of motion due to the inertial oscillations: positive Fourier frequencies correspond to counter-clockwise (CCW) rotation and negative frequencies to clockwise (CW) rotation; in the Northern Hemisphere, inertial motion contributes a CW enhancement near the local Coriolis frequency, while in the Southern Hemisphere it contributes a CCW enhancement under the same convention [42]. Figures presenting the results of the analysis therefore include a vertical line at the Coriolis frequency computed from the mean deployment latitude (Table 2). The principal lunar semidiurnal constituent $M_2$ has nominal frequency $f_{\mathrm{M2}} = 1.9322$ cpd; near the Coriolis band, inertial and tidal signatures are not fully separable at the present sampling and record lengths when enhancements are present in both the CW and CCW spectra, so secondary peaks in that neighbourhood are interpreted cautiously: enhancement confined mainly to the CW spectrum is taken as evidence of inertial influence, whereas enhancement in both the CW and CCW spectra indicates a more elliptical response consistent with additional tidal or internal-stress influences rather than a pure inertial line.

To smooth the spectral estimates for plotting and comparison, rotary spectra are averaged into logarithmically spaced frequency bins from $f_{\min}$ up to a nominal display cut-off of 6 cpd. This cut-off is below half the Nyquist frequency of the hourly Arctic composites and excludes the shortest-period end of the higher-resolution Antarctic records. The 3 hour SIPEX II Waves campaign is the exception, with usable frequency support ending at its lower Nyquist limit of 4 cpd. Bin edges $f_j$ for $j = 0, ..., N_b$ are geometrically spaced with $N_b = 50$; within each bin the frequency coordinate is taken as the geometric mean of the edges and the spectral values as arithmetic means across contributing Fourier bins. The choice of 50 bins was made from the processed record lengths and sampling intervals.[2] Across the 2017-2024 Arctic buoy records this setting retains a median of 36 displayed bins up to 6 cpd, while the Antarctic campaign records retain a median of 29 displayed bins; even the shortest SHARC 2022 records retain 14-17 displayed bins. The resulting bin spacing is typically about 14% in frequency, which reduces raw periodogram variance while preserving the low-frequency slope and the near-Coriolis band used in the interpretation. Integrated CW variance and log–log spectral slopes reported later are evaluated over the same binned grids so that energy and slope summaries remain consistent with the figures.

[1] A window-sensitivity analysis recomputing the rotary spectra that follow with Hann, Blackman, rectangular, and Tukey ($\alpha = 0.25$) tapers while holding the preprocessing, PSD normalisation, and log-binning fixed showed that the main spectral-shape diagnostics were stable for nearby moderate tapers: relative to the Hamming reference, Hann changed the median branch-wise RMS log-PSD by only 0.058 and the median total resolved power by 5.9% across the sampled buoy records, whereas more dissimilar tapers mainly affected absolute PSD level and band-integrated variance; this supports the use of the Hamming window here as a compromise between leakage suppression and frequency resolution.

### 3.1.3 Principal component analysis

PCA is then applied to common-time matrices of buoy velocity components within each Arctic regime–season composite and each Antarctic campaign. Each physical buoy contributes two columns, the eastward and northward components $(u, v)$, so that the PCA acts on the vector drift field. Rows correspond to time steps and columns to the $M$ retained component series. Columns are first standardised to zero mean and unit variance. To account for heavy-tailed, non-Gaussian drift variability, the correlation matrix $\boldsymbol{R}$ is constructed from Spearman rank correlations rather than Pearson correlations. The covariance matrix is then reconstructed from $\boldsymbol{R}$ and the original standard deviations. Letting $\boldsymbol{D} = \mathrm{diag}(\sigma_1, ..., \sigma_M)$ denote the diagonal matrix of the $M$ component standard deviations, we have

$$\boldsymbol{\Sigma} = \boldsymbol{D}\,\boldsymbol{R}\,\boldsymbol{D}. \tag{8}$$

Eigenvectors $\boldsymbol{v}_k$ and eigenvalues $\lambda_k$ satisfy the eigenproblem

$$\boldsymbol{\Sigma}\boldsymbol{v}_k = \lambda_k \boldsymbol{v}_k, \tag{9}$$

with eigenvalues ordered so that $\lambda_1 \geq \lambda_2 \geq \cdots \geq \lambda_M$. The $k$th principal component time series is defined by

$$\mathrm{PC}_k(t) = \sum_{i=1}^{M} v_{ki}\, x_i(t), \tag{10}$$

where $x_i(t)$ denotes one of the retained velocity-component series. The fraction of variance captured by the first $K$ modes is

$$\frac{\sum_{k=1}^{K} \lambda_k}{\sum_{k=1}^{M} \lambda_k}. \tag{11}$$

PCA is used descriptively: high cumulative variance in the leading components indicates that much of the coherent array-scale vector motion projects onto a low-dimensional subspace, but it does not by itself establish the importance of deformation, waves, or local forcing at higher frequencies. For Arctic records, PCA is run on a contiguous window where enough buoys overlap; for Antarctic notebooks, only timestamps with valid data on all retained velocity components are kept. To show which frequencies each mode carries, the same Hamming-windowed spectral approach used for buoy spectra is applied to each $\mathrm{PC}_k(t)$. These are spectra of the modal score time series, not of individual buoy velocities, so their absolute levels should be interpreted as array-scale modal variance rather than one-to-one equivalents of single-buoy rotary PSD.

[2] A sensitivity analysis repeating the shape-distance and amplitude-sensitive root mean square difference calculations (described in Section 3.2.1) with 35 and 70 log-spaced bins showed stable pairwise rankings relative to the 50-bin reference across all 45 spectrum pairs, with Spearman correlations of 0.956-0.965 for shape distance and 0.979-0.982 for amplitude-sensitive RMSD; this stability supports the use of 50 bins as a defensible compromise between smoothing and spectral resolution.

### 3.1.4 Uncertainty quantification

Uncertainty bands throughout the spectra presented below represent ensemble spread in the observed drift field rather than propagated positional error. GNSS positioning uncertainty contributes most directly to the highest-frequency velocity estimates. Using the 10 m position proxy noted above and assuming independent errors between consecutive fixes gives an approximate step uncertainty of $\sqrt{2} \times$ 10 m. Expressed as an equivalent velocity uncertainty, this is appropriately 1.6 cm/s at 15 min sampling, 1.2 cm/s at 20 min, 0.4 cm/s at 1 h, 0.13 cm/s at 3 h, and 0.07 cm/s at 6 h. These values are most relevant to the short-period end of the spectra and to weak high-frequency variance; the main comparisons below emphasise low-frequency and near-inertial structure where ensemble spread, array geometry, and record length dominate the uncertainty.

For rotary spectra, both the Arctic regime composites and the Antarctic campaigns are summarised on common log-frequency grids by the ensemble median with an inter-quartile range (IQR) (25–75th percentile) across the contributing spectra. The only partial exception is the Antarctic campaign PCA summaries: because a campaign-level principal component is defined for the full array rather than for independent per-buoy spectra, spread around the leading PC spectra is estimated by recomputing the PCA after omitting each buoy in turn and then taking the inter-quartile range across those leave-one-buoy-out spectra. This leave-one-buoy-out construction is therefore used as the Antarctic PCA analogue of ensemble spread, not as a formal measurement-error confidence interval.

## 3.2 Cross-polar spectral comparison

The inter-regional comparison then uses these regime-season and campaign-level products to compare Arctic and Antarctic behaviour in terms of spectral scale, spectral shape, rotary asymmetry, coherence, and modal compactness.

### 3.2.1 Spectral structure and energy across regions

First, the windowed spectral scale of the spectra is summarised by integrating CW, CCW, and total resolved PSD over 0.1–6 cpd where supported by the sampling grid. Spectral shape is then compared separately using variance-normalised band fractions, sub-inertial slopes, and distances between area-normalised log spectra, so amplitude similarity is not conflated with similarity of spectral envelope. Rotary asymmetry is quantified in a fixed window around the local Coriolis frequency, denoted here by $f_c$, using the fraction of near-$f_c$ variance carried by the hemisphere-appropriate inertial branch.

Cross-regional spectral comparison is then presented using two complementary distance metrics: one sensitive to spectral shape only, and one sensitive to absolute amplitude. Both metrics operate on the median dominant-branch rotary spectra evaluated on a shared log-spaced frequency grid spanning 0.1–6 cpd where supported by the underlying sampling. Here, $S_{\mathrm{dom}}(f)$ denotes the regime-season or campaign-level median of the spectra sensitive to inertial forcing: CW in the Arctic and CCW in the Antarctic. This hemisphere-appropriate selection preserves physically like-for-like comparison across polar regions. Because some campaigns do not populate the full grid, both distance metrics are evaluated only on the subset of frequency bins valid for all spectra in the comparison set, ensuring every pairwise value is computed over identical frequency support. The paired use of these metrics therefore separates similarity in variance distribution from similarity in absolute spectral energy level.

#### 3.2.1.1 Area-normalised spectral shape

For the shape-only comparison each median dominant-branch spectrum $S_{\mathrm{dom}}(f)$ is evaluated on the common log-spaced frequency grid (0.1–6 cpd) and area-normalised by a trapezoidal integral $A$ over that band, yielding the normalised spectrum $S_{\mathrm{n}}(f)$,

$$A = \int S_{\mathrm{dom}}(f)\,\mathrm{d}f, \quad S_{\mathrm{n}}(f) = \frac{S_{\mathrm{dom}}(f)}{A}, \tag{12}$$

so that differences in overall spectral level are removed before comparison. The distance between two campaigns is then defined as the root-mean-square difference (RMSD) of $\log_{10} S_{\mathrm{n}}(f)$ across the shared frequency grid. This metric therefore compares the spectral envelope only, emphasising relative peaks, troughs, and slopes across frequency rather than absolute variance.

#### 3.2.1.2 Absolute spectral amplitude differences

For the amplitude-sensitive comparison, no area normalisation is applied and the spectral separation between two campaigns is instead taken as the RMSD of the raw dominant-branch rotary PSDs across the same common frequency grid,

$$\mathrm{RMSD}_{ij} = \sqrt{\left(\frac{1}{N}\right) \sum_{k=1}^{N} \left(S_{\mathrm{dom},i}(f_k) - S_{\mathrm{dom},j}(f_k)\right)^2}, \tag{13}$$

where $S_{\mathrm{dom},i}(f_k)$ and $S_{\mathrm{dom},j}(f_k)$ are the dominant-branch spectra of campaigns $i$ and $j$ evaluated at the $k$th frequency on the shared grid and $N$ is the number of valid frequency bins. Reported values are expressed in $(\mathrm{cm/s})^2/\mathrm{cpd}$, so this metric retains differences in both spectral amplitude and spectral structure. It is a difference metric between observed spectra, not a prediction error.

### 3.2.2 Inter-buoy coherence across regions

Linear coupling between buoy pairs as a function of frequency is summarised by magnitude-squared coherence of the velocity components, computed with Welch's averaged periodogram (50% overlap, Hamming-windowed segments) over the temporal overlap shared by the pair after interpolation onto a common time grid and removal of the mean from each series. For the eastward and northward components separately, the coherence is determined by,

$$C_u(f) = \frac{\left|P_{u_1 u_2}(f)\right|^2}{P_{u_1 u_1}(f)\, P_{u_2 u_2}(f)}, \quad C_v(f) = \frac{\left|P_{v_1 v_2}(f)\right|^2}{P_{v_1 v_1}(f)\, P_{v_2 v_2}(f)}, \tag{14}$$

where $P_{xy}(f)$ denotes the Welch-estimated cross-spectrum and $P_{xx}(f)$ the corresponding auto-spectrum. The single coherence curve reported for each buoy pair is then

$$C_{xy}(f) = \frac{C_u(f) + C_v(f)}{2}, \quad 0 \leq C_{xy}(f) \leq 1. \tag{15}$$

Values near unity indicate correlated velocity variability at that frequency (shared advection or large-scale forcing), whilst values near zero indicate largely independent motion. Inter-buoy coherence is most informative at synoptic and sub-inertial scales, where array-scale organisation is expected, and typically weakens toward higher frequencies where local ice mechanics and noise dominate.

### 3.2.3 Cross-regional PCA mode structure

Lastly, following from the PCA analysis in each regime-season and campaign, now the spectra of the leading principal components are compared across regions to assess whether the dominant modes occupy similar broad frequency bands, especially at low frequency and near the local Coriolis band. This comparison is interpreted as a measure of modal compactness and shared timescale structure within arrays, not as direct evidence that the underlying dynamics or force balances are the same between the Arctic and SO.

# 4 Sea Ice Drift in the Arctic

## 4.1 Regimes and seasonal variability (BG vs TD)

As stated in the data description, the Arctic is divided into two regimes according to the dominant currents, the BG and the TD Stream. The BG forces anticyclonic, CW circulation in the Amerasian

Arctic study area and drift tracks

Sample GPS buoy drift tracks (Spring 2021) and regional boundaries for Beaufort Gyre (BG) and Transpolar Drift (TD).

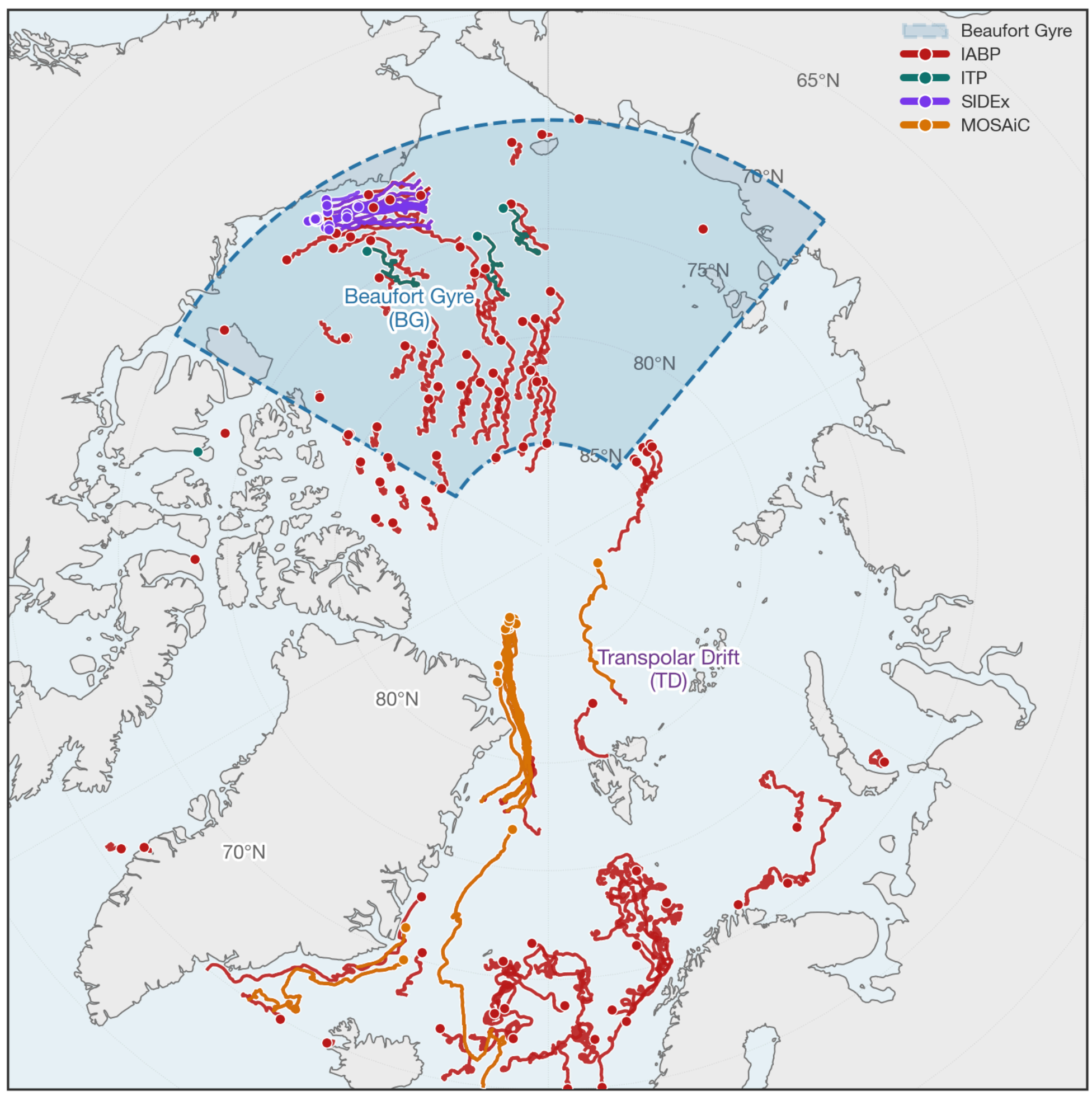

**Fig. 1** Arctic study area showing example Spring 2021 drift tracks from IABP, ITP, SIDEx, and MOSAiC. The dashed blue polygon marks the BG analysis domain; the TD label denotes the remaining Arctic study domain north of 65°N outside BG. NOMAD drift tracks are not visible because that deployment only took place in 2022.

Basin, while the TD Stream carries ice from the eastern and central Arctic into the North Atlantic through the Fram Strait. For the purpose of this study, the BG region is defined as a domain bounded by 70-85°N and 140°E-120°W, and is based on the BGx definition in [43]. The TD region is defined, broadly, as all positions north of 65°N falling outside the BG region. This is shown in Fig. 1 above.

Table 2 that follows, lists summarising statistics of these Arctic data. Here it can be seen that TD data consistently contain more buoys than BG in every season (approximately 113–161 buoys for TD versus 49–67 for BG), whilst mean deployment durations remain broadly comparable at 61–76 days. Consequently, the principal structural distinction between these regimes is in the spatial organisation and data extent rather than in record length. The Arctic reference spectra presented herein should therefore be interpreted as regime-specific climatologies, ensuring that subsequent comparison with Antarctic campaigns is conducted against two Arctic baselines rather than against a single, generic Arctic mean.

**Table 2** BG/TD seasonal statistics: number of datasets, mean duration, buoy count, spatial bounds.

| Regime | Season | Datasets | Mean buoy count | Mean duration (days) | Latitude range (°N) | Longitude range (°) |
|---|---|---|---|---|---|---|
| BG | Winter | 8 | 48.50 | 64.04 | 67.0 to 84.8 | −179.7 to −120.2 |
| BG | Spring | 8 | 57.25 | 75.54 | 67.0 to 84.7 | −179.9 to −120.2 |
| BG | Summer | 8 | 59.88 | 65.11 | 65.5 to 84.9 | −179.8 to −120.2 |
| BG | Autumn | 8 | 66.62 | 63.00 | 66.9 to 85.0 | −179.9 to −120.1 |
| TD | Winter | 8 | 129.12 | 67.83 | 65.0 to 89.0 | −165.8 to 179.9 |
| TD | Spring | 8 | 112.62 | 72.84 | 65.0 to 89.7 | −167.6 to 180.0 |
| TD | Summer | 8 | 130.38 | 60.76 | 65.0 to 89.9 | −166.8 to 179.6 |
| TD | Autumn | 8 | 161.00 | 68.65 | 65.0 to 89.3 | −171.0 to 179.9 |

## 4.2 Rotary spectra of the Arctic regimes

From the regime and seasonally based data, the log-binned rotary spectra for the Arctic were then computed. Fig. 2 below shows the rotary spectra for the BG regime, whilst Fig. 3 that follows shows the rotary spectra for the TD regime.

For both Arctic regimes it is clear that drift in the Arctic is dominated by low-frequency variability, with the maximum amplitudes approaching $10^{-2}$ $(\mathrm{m/s})^2$/cpd in the PSD occurring at frequencies around 0.1 cpd and with power decreasing with increasing frequency. Secondary enhancements are clear around the Coriolis frequency point in the CW spectra for the BG and TD regimes with amplitudes ranging between $10^{-4}$–$10^{-2}$ $(\mathrm{m/s})^2$/cpd, but which are not present in the median CCW spectra for the BG regime and are weaker than the corresponding CW spectra for the TD regime. This suggests that, in addition to primary low-frequency variability in both regimes, drift in the BG regime is susceptible to inertial oscillations (peaks in only the CW spectra around the Coriolis frequency), while the TD regime shows a less one-sided inertial-semidiurnal band with peaks in both CW and CCW spectra. This indicates that the TD regime's near-Coriolis frequency band cannot be cleanly attributed to only inertial oscillation but reflects the influence of other drivers in this frequency range such as tides and internal stresses.

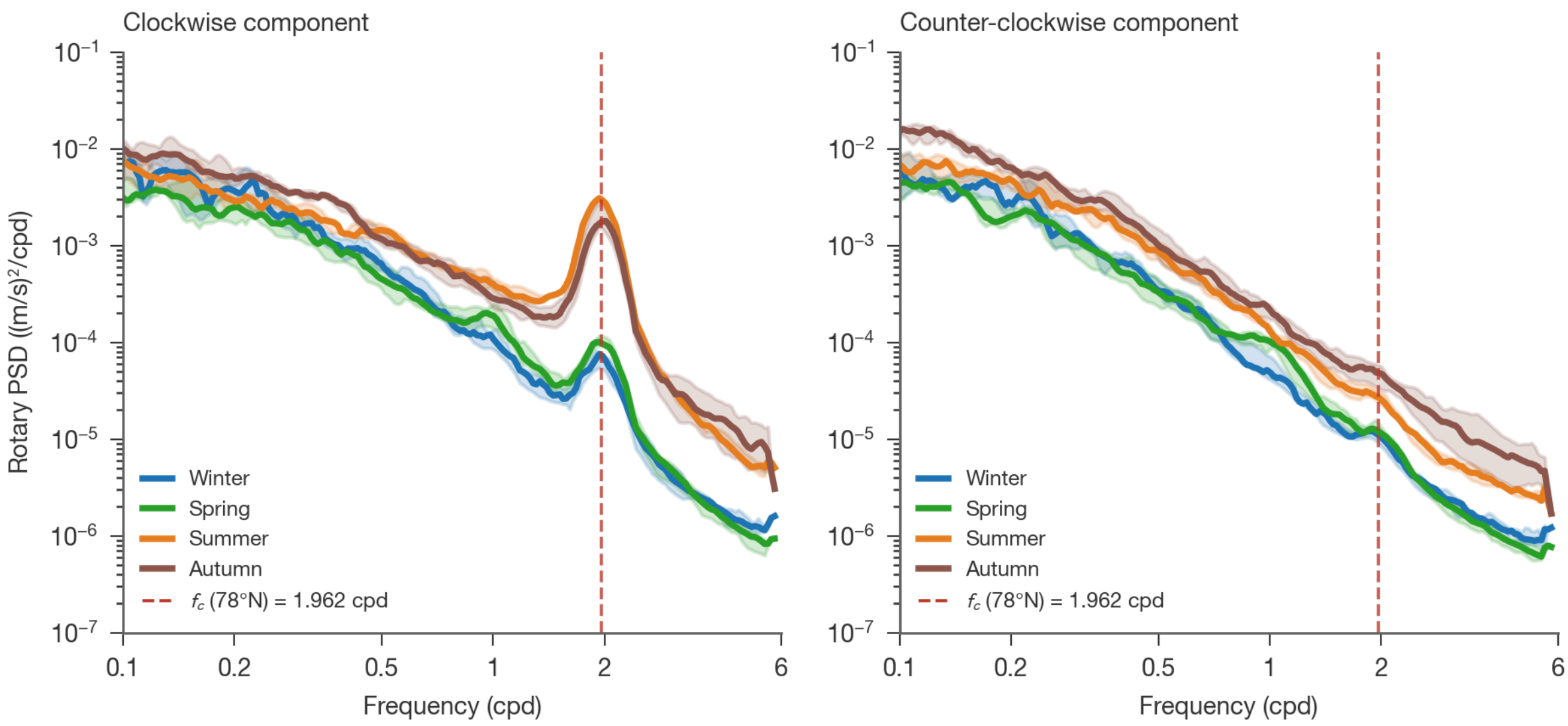

**Fig. 2** Seasonal rotary spectra in the BG region, 2017–2024. Lines show seasonal median CW and CCW rotary PSD; shaded bands show the IQR across contributing regime-season spectra, and vertical dashed lines mark the representative Coriolis frequency.

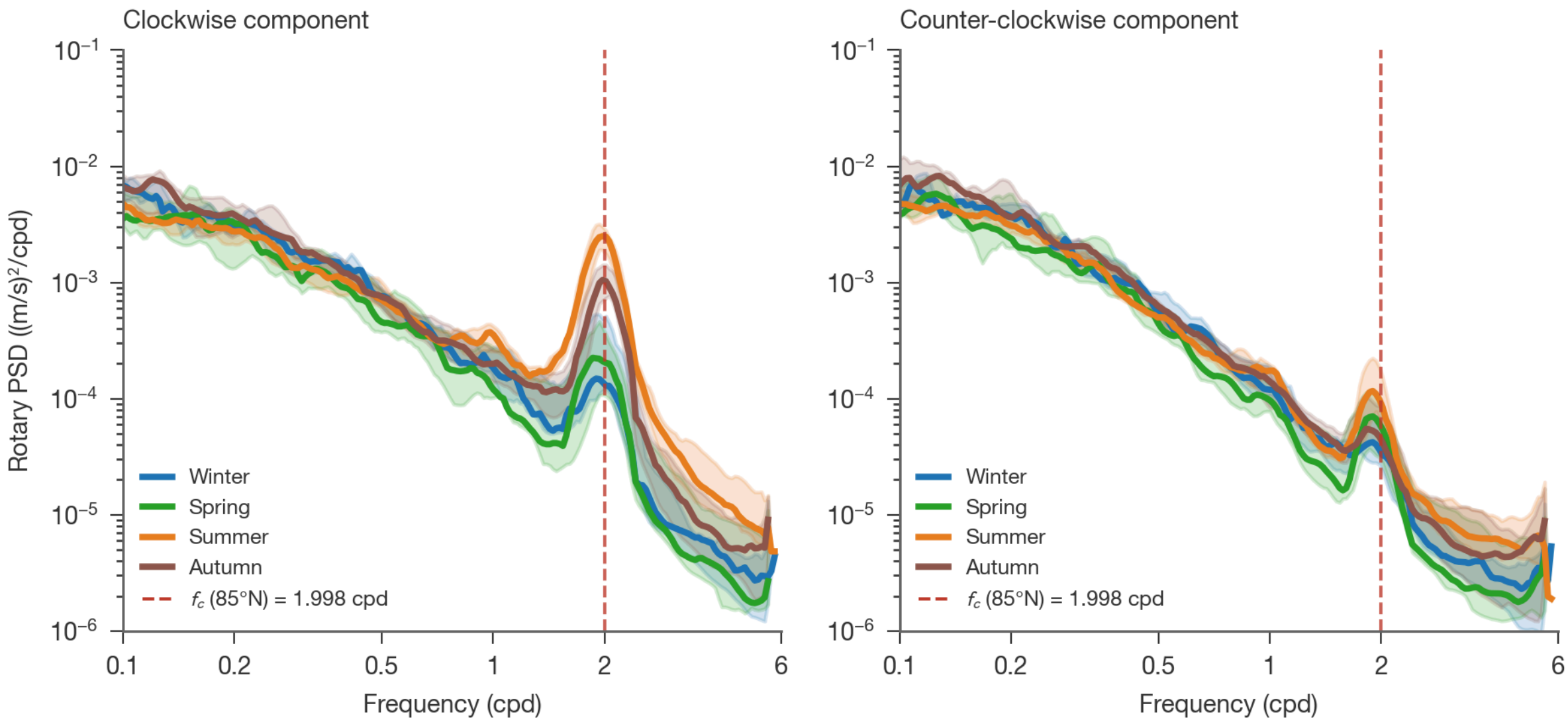

**Fig. 3** Seasonal rotary spectra in the TD region, 2017–2024. Lines show seasonal median CW and CCW rotary PSD; shaded bands show the IQR across contributing regime-season spectra, and vertical dashed lines mark the representative Coriolis frequency.

Furthermore, it is clear from the greater magnitude of these peaks in the Summer and Autumn seasons in both regimes that drift in these seasons is more energetic than in the Winter and Spring seasons, although the differences are more pronounced in the BG regime. This seasonal contrast is consistent with reduced ice consolidation and stronger wind-ocean coupling in summer and autumn, although those mechanisms are not diagnosed directly from these spectra.

Taken together, Fig. 2 and Fig. 3 show that ice drift in the BG and TD regimes share the same broad low-frequency spectral structure, but ice in the TD regime exhibits a more evident CCW response near the local Coriolis frequency and slightly elevated variance above $\approx 2$ cpd, indicating a somewhat less asymmetric rotary response than in the BG regime.

## 4.3 Principal component analysis of the Arctic regimes

PCA was then applied to the Arctic velocity data to further investigate the structure of the Arctic drift. Fig. 4 below shows the PCA power spectra for the BG and TD regimes.

These spectra describe the frequency content of the PC score time series, so they summarise coherent modal variance of the buoy array rather than the PSD of a representative individual buoy.

The shapes of the spectral curves for the first two principal components (PCs) in the BG region are similar to each other and operate at similar magnitudes in corresponding frequency bands with both dominated by low frequency variability and a weaker enhancement near the local Coriolis frequency. This indicates that the leading modes of drift in the BG region represent different expressions of common gyre-scale variability rather than distinct timescales.

In the lower row of Fig. 4, the PCA power spectra for the first two modes in the TD region are shown. These are less similar to each other than in the BG region and operate at scales of approximately one order of magnitude larger than in the BG region. However, they follow a similar broad pattern of low-frequency dominance and a weaker enhancement near the local Coriolis frequency. PC1 is more strongly weighted towards the lowest frequencies, and the near-Coriolis enhancement is broader and shifted leftwards to lower frequencies. In contrast, PC2 is relatively clearer, the Coriolis enhancement peaks at the local Coriolis frequency and is narrower. Further, PC1 shows a much larger IQR than PC2 (or either PC in the BG regime), which suggests greater data set-to-data set variability in the TD region than in the BG region.

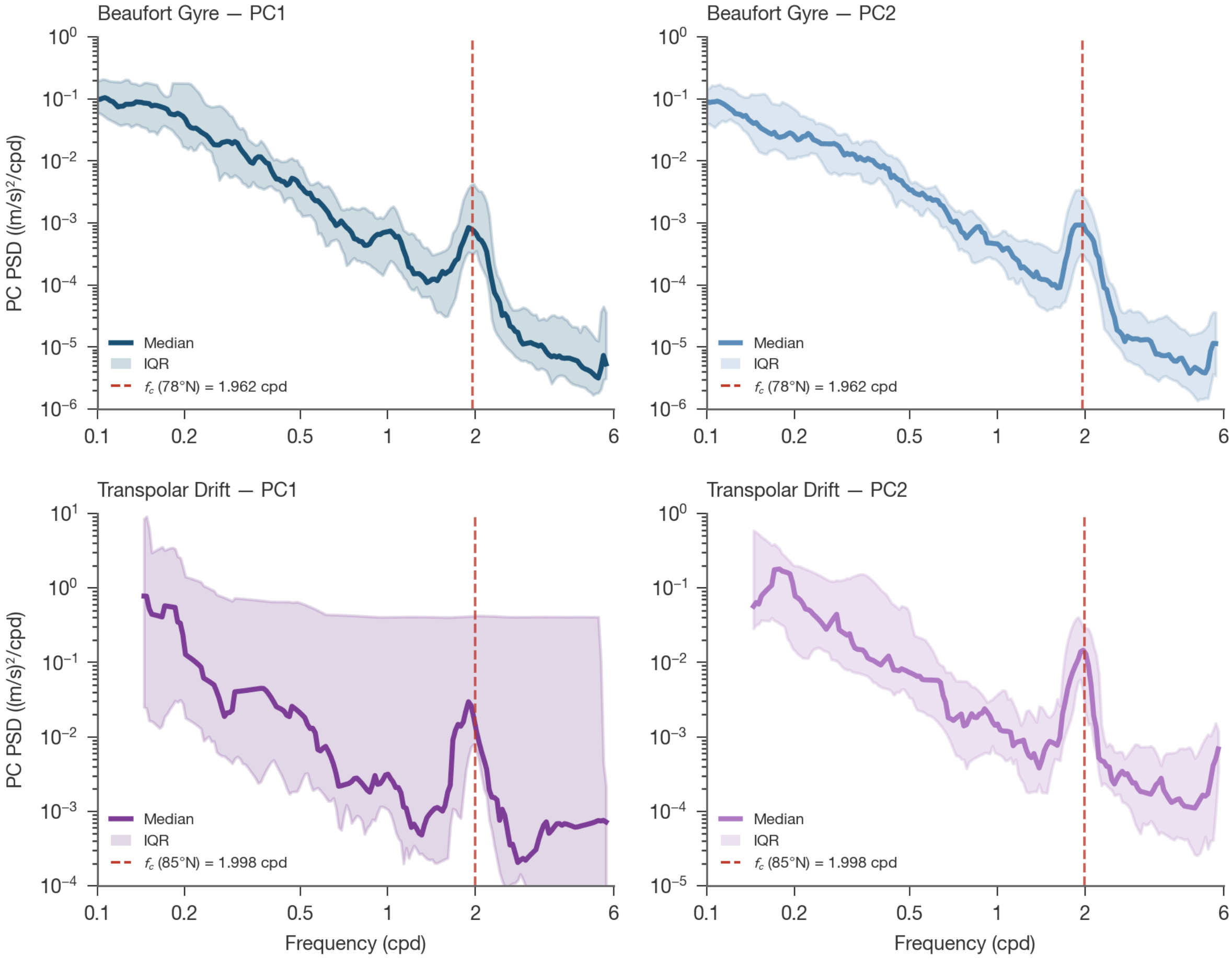

**Fig. 4** PCA score power spectra for the BG and TD regions. In each row, panels show PC1 and PC2 median spectra across regime-season datasets; shaded bands show the IQR, and the vertical dashed line marks the representative Coriolis frequency. The upper row shows the BG regime and the lower row shows the TD regime.

Overall, these results suggest that the BG and TD regimes share similar leading-mode frequency content, while the TD regime shows a less compact and less repeatable modal representation than the BG regime, consistent with its larger spatial extent and stronger pathway variability.

Table 3 below, presents the mean variance explained by the first three PCs for the BG and TD regimes by season. Here it can be seen that in both regimes, across all seasons, that PC1 is consistently the dominant mode accounting for between 30% and 48% of the total variance in the data. PC1 also always accounts for more than double the variance of PC2. PC2 dominates PC3 by a relatively smaller margin and these modes capture a similar amount of total variance in the data. This suggests a dominant leading mode for ice motion across the Arctic with higher order variability spread across several weaker modes. This supports the idea that, over the sampled regime-season windows, coherent Arctic array-scale drift has relatively low dimensionality, with a dominant leading mode and a few weaker modes carrying the remaining variability.

**Table 3** Mean variance explained by the first three PCs for BG/TD by season.

| Regime | Season | Mean PC1 % variance | Mean PC2 % variance | Mean PC3 % variance | Cumulative PC1 to PC3 % variance |
|---|---|---|---|---|---|
| BG | Winter | 36.9 | 19.0 | 12.5 | 68.4 |
| BG | Spring | 48.0 | 17.2 | 9.1 | 74.4 |
| BG | Summer | 34.8 | 13.1 | 8.1 | 56.0 |
| BG | Autumn | 30.0 | 14.9 | 10.3 | 55.1 |
| TD | Winter | 34.1 | 13.7 | 11.0 | 58.8 |
| TD | Spring | 39.4 | 10.3 | 7.9 | 57.5 |
| TD | Summer | 33.6 | 12.4 | 10.1 | 56.0 |
| TD | Autumn | 39.7 | 10.7 | 7.3 | 57.7 |

## 5 Sea Ice Drift in the Southern Ocean

### 5.1 Southern Ocean campaigns

These analyses for the Arctic were then repeated for the SO in the sections that follow. These campaigns are shown relative to the Antarctic continent in Fig. 5 below, and are summarised in Table 4 of Section 5.1. As noted earlier, the SO is extremely data sparse for in-situ sea ice drift data in comparison to the Arctic and so these data sets are presented as campaign specific results rather than ensemble by region and season.

Fig. 5 and Table 4 show clearly that the Antarctic campaigns are heterogeneous in terms of location, spread, number of buoys and length. In general, campaigns are comprised of low buoy numbers (2–8) and short experiment length over small geographical areas. This is largely due to the difficulty of accessing the environment and the harsh conditions in which buoys are required to survive. A further difference between the SO campaign analysis and the Arctic analysis above is that the local Coriolis frequency is calculated per campaign due to the significant variation in mean latitude between campaign. Consequently, the inertial band is not fixed across the Antarctic cases and must be interpreted locally.

### 5.2 Rotary spectral analysis of the SO campaigns

As before, first, the rotary spectra for the experiments were calculated. Due to the lower number of buoys and the relatively short length of the Antarctic data sets, spread is summarised directly from the ensemble of buoy spectra within each campaign using the IQR rather than from a pooled climatological distribution. Fig. 6, overleaf, shows the rotary spectra for the SO data sets.

Here, it can be seen that there are wide variations between the SO campaigns. However, as with the Arctic, all spectra show energy dominance in the low frequencies with decreasing power with increasing frequency. Here, a CCW (purple spectra) enhancement is evident in most regions around the local Coriolis frequency. In conjunction, most regions do not show this enhancement at the corresponding frequencies in the CW spectra. This is consistent with the expected Southern Hemisphere rotary sense of inertial motion. Notably, this pattern is not clear in the SIPEX II Buoys data from 2012, where the CW spectra also show an uptick in the frequencies near the local Coriolis frequency. As before, with the TD regime in the Arctic, this indicates a more mixed inertial-semidiurnal band rather than a uniquely inertial signal. The SCALE Winter 2019 data is also an outlier in this respect with the appearance of a peak in the CW spectra as well around the 2 cpd frequency. Notably here we see some separation in frequency between

Antarctic study area and campaign drift tracks

GPS buoy drift trajectories from eight observational campaigns across the Weddell, Ross, and East Antarctic seas.

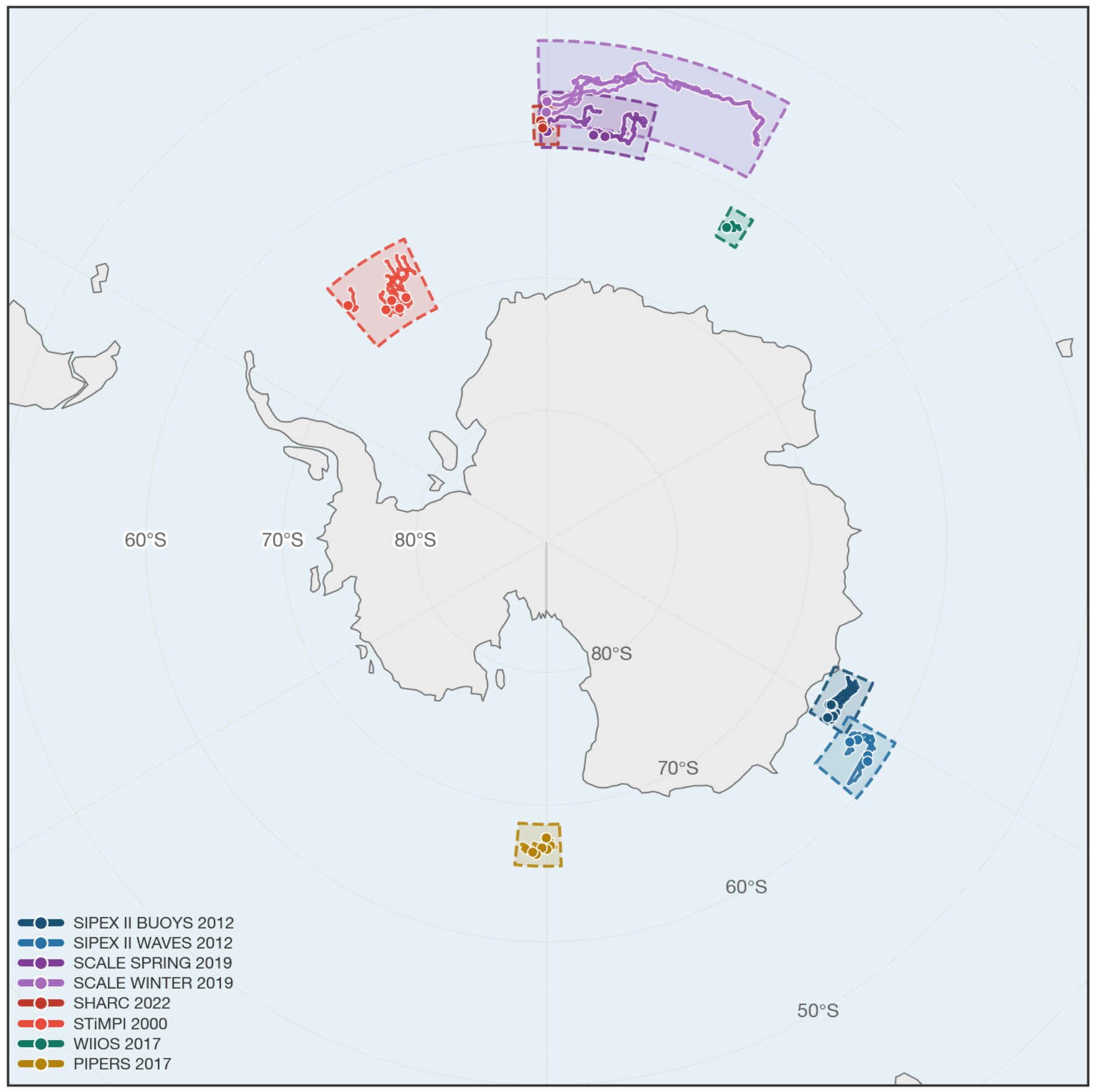

**Fig. 5** Antarctic study area showing campaign drift tracks between 55°S and 90°S. Colours distinguish campaigns, filled circles mark track starts, and the map extent shows the deployments relative to the Antarctic continent.

the CCW and CW peaks with the CCW enhancement shifted left of the local Coriolis frequency and the CW enhancement shifted towards higher frequencies.

Notably, with the exception of the SIPEX II Buoys 2012 and SHARC 2022 campaigns, the power spectra shown in Fig. 6 are approximately one order of magnitude larger in the SO than in the Arctic with low-frequency power usually approaching $10^{-1}$ $(\text{m/s})^2/\text{cpd}$. As can be seen in Fig. 5 on the previous page, SIPEX II Buoys 2012 was closer to the Antarctic continent than the other campaigns, so its lower spectral amplitude may plausibly reflect more consolidated or mechanically constrained ice, although this interpretation is not independently tested here. The SHARC 2022 experiment is the shortest experiment capturing drift over just 1.7 days and so should be considered an energy outlier in this analysis.

The influence of inertial oscillations in some regions (SIPEX II Waves 2012, WIIOS 2017, SCALE SPRING 2019) is also more significant than in the Arctic, with CCW Coriolis enhancements being up to 3 orders of magnitude larger than the corresponding amplitude in the CW spectra. In the Arctic this difference only approached a single order of magnitude. In general, the SO drift spectra are organised similarly in structure to the Arctic from a shape and frequency perspective. The higher SO energy level is consistent with less consolidated ice, stronger wind and wave exposure, and a greater role for ocean and ice-edge processes.

**Table 4** Overview of the Antarctic campaigns on the common overlapping analysis window. For each campaign the table lists the deployment year, the number of buoys, the mean duration in days, the mean latitude and longitude, the corresponding Coriolis frequency, and the start and end dates.

| Campaign | Year | Buoy count | Mean duration (days) | Mean latitude (°) | Mean longitude (°) | Coriolis frequency (cpd) | Start date | End date |
|---|---|---|---|---|---|---|---|---|
| SIPEX II Buoys 2012 | 2012 | 8 | 17.5000 | −64.6230 | 116.4980 | 1.8120 | 2012-10-23 | 2012-11-09 |
| SIPEX II Waves 2012 | 2012 | 4 | 8.1250 | −61.8524 | 122.4657 | 1.7683 | 2012-09-24 | 2012-10-02 |
| SCALE Spring 2019 | 2019 | 3 | 50.8750 | −58.3352 | 8.9200 | 1.7069 | 2019-10-29 | 2019-12-19 |
| SCALE Winter 2019 | 2019 | 3 | 27.9167 | −56.3426 | 4.4255 | 1.6693 | 2019-07-28 | 2019-08-25 |
| SHARC 2022 | 2022 | 3 | 1.7396 | −58.7784 | −0.4787 | 1.7150 | 2022-07-20 | 2022-07-22 |
| STiMPI 2000 | 2000 | 6 | 17.9444 | −68.4479 | −33.3516 | 1.8653 | 2000-04-20 | 2000-05-08 |
| WIIOS 2017 | 2017 | 2 | 8.7604 | −62.4734 | 30.7416 | 1.7784 | 2017-07-04 | 2017-07-13 |
| PIPERS 2017 | 2017 | 6 | 8.8438 | −66.9495 | 181.2905 | 1.8454 | 2017-06-13 | 2017-06-22 |

### 5.3 Principal component analysis of the SO campaigns

Continuing from the spectral analysis, PCA was again applied to the SO data sets. Fig. 7 below shows the spectral composition of the first two modes of these data. Across most campaigns, PC1 and PC2 are dominated by variance below about 0.5 cpd, with spectral levels typically decreasing by roughly one to three orders of magnitude from the lowest resolved frequencies to the high-frequency end of the band. Several campaigns also show a secondary enhancement near the local Coriolis frequency, which varies between approximately 1.66 and 1.86 cpd across the deployments, although the strength and clarity of that peak differ substantially between campaigns; these are most clear in the SCALE Spring 2019 spectra and imperceptible in SIPEX II Waves 2012 and SHARC 2022.

Table 5 below summarises the cumulative variance captured by the first two modes for the SO campaigns. Here it is clear that the first two modes account for more than 79% of the variance in each SO campaign and so the data sets are significantly lower dimensional than the Arctic composites. These high values confirm a strong shared synoptic-scale drift signal within each array, leading to the result that Antarctic arrays often contain a dominant common mode. However, very small arrays naturally inflate the variance captured by the first two components, so these high values are not necessarily predictive of large ensembles or of longer time scales than reflected by the data. Given this, PCA supports efficient capture of the leading common drift mode, not the irrelevance of dense arrays for different experimental objectives.

## 6 Cross-Polar Spectral Comparison

Following the regional level analysis, this section now compares drift in the Arctic to the SO using the methodology outlined in Section 3.2.

### 6.1 Spectral structure and energy across regions

Table 6 overleaf presents an inter-regional comparison summary by comparing the variance contained in the rotary spectra of Section 4 and Section 5. The proportions of variance are reported for the overall comparison and by frequency band: low frequencies (below 0.8 cpd), the band around $f_c \pm 0.25$ cpd, and frequencies above 2 cpd.

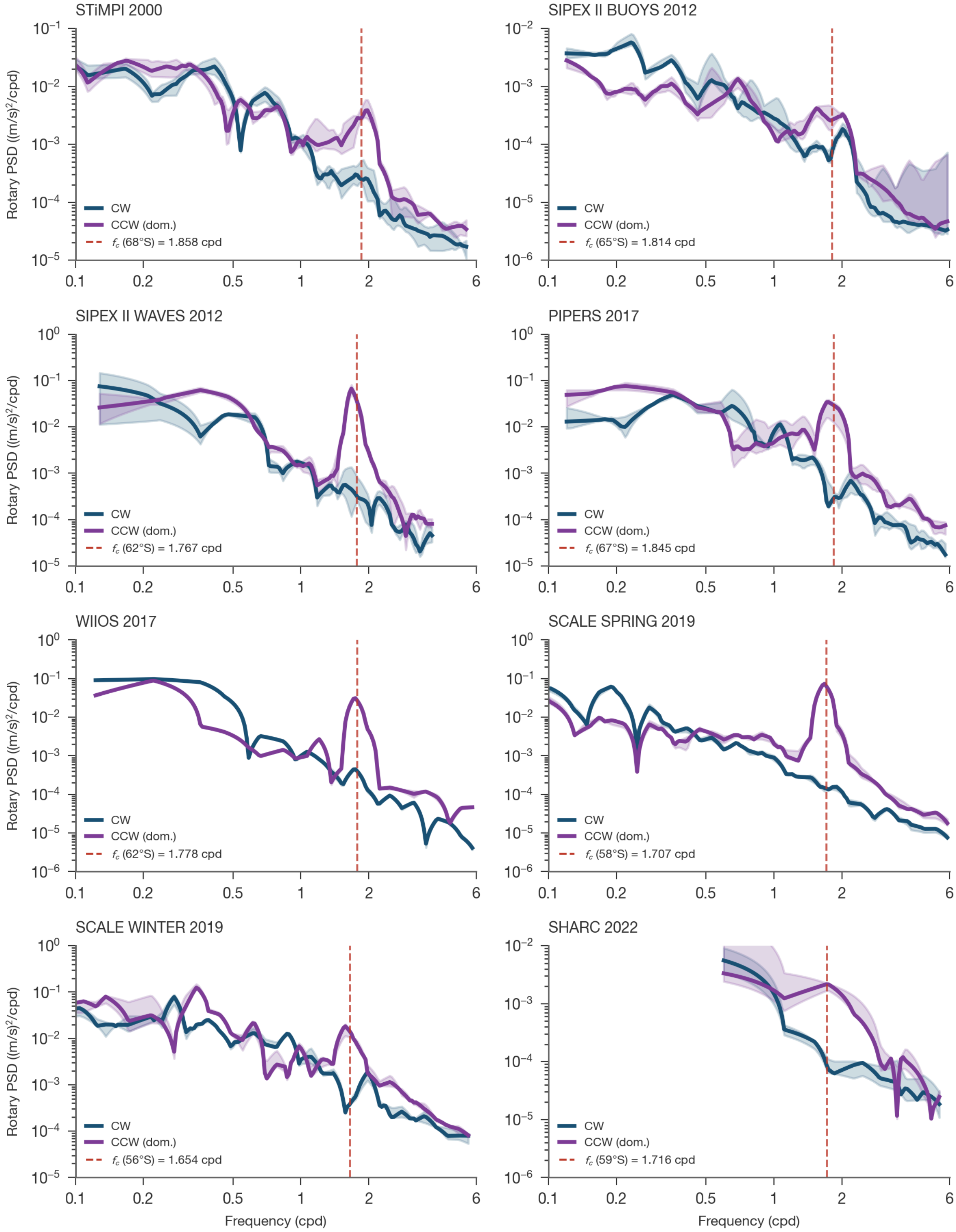

**Fig. 6** Rotary spectra for the Antarctic campaigns are shown in eight panels arranged four-by-two, ordered from the oldest campaign (2000) to the newest (2022). Within each panel, the median clockwise and counter-clockwise spectra are shown together with the inter-buoy interquartile range, and the local Coriolis frequency is indicated at the mean latitude of the campaign. The y-axis limits vary independently between panels so that the internal structure of each campaign remains visible; cross-campaign amplitude comparisons are therefore deferred to Table 6 and Fig. 9.

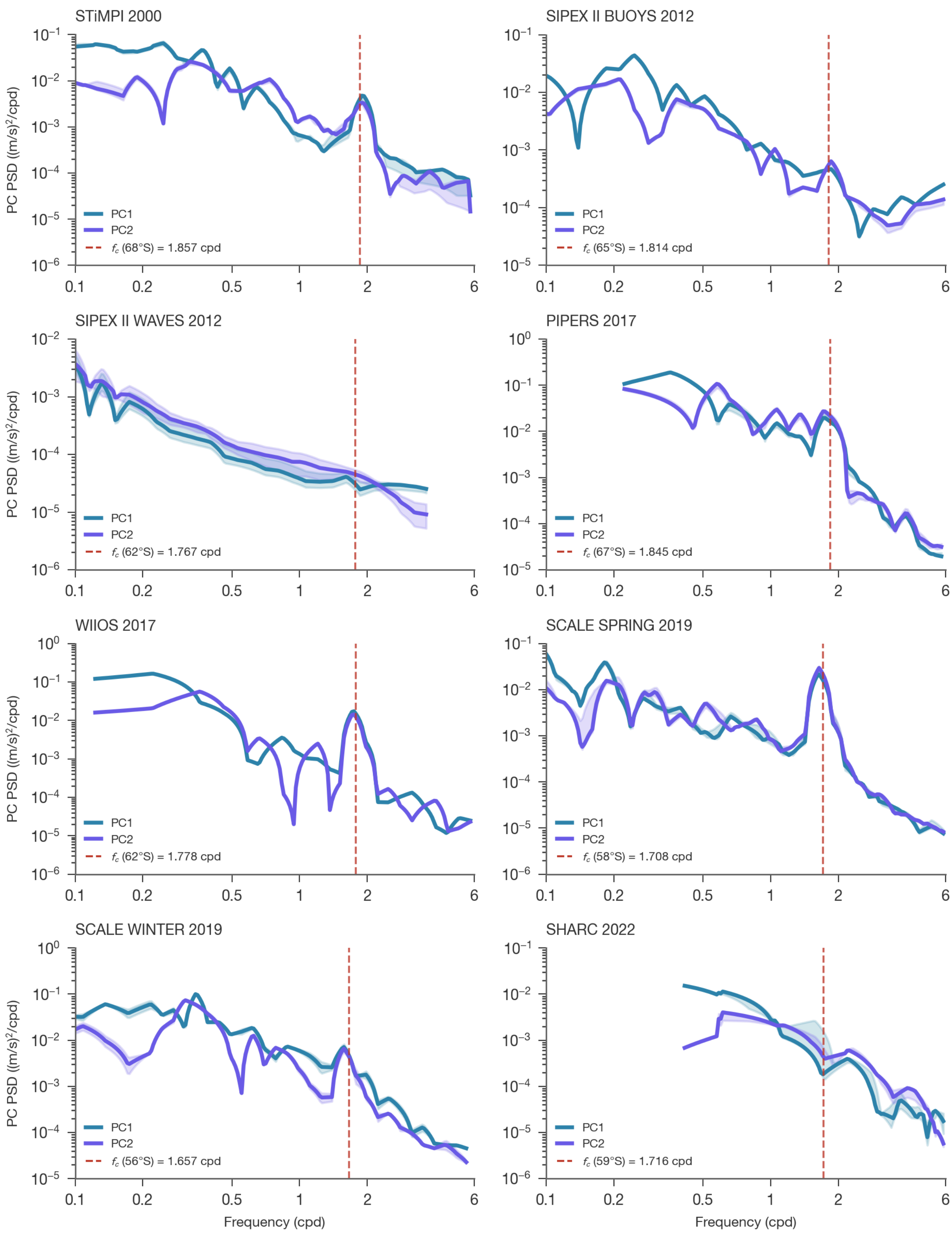

**Fig. 7** Antarctic PCA score power spectra by campaign. One panel per campaign, ordered oldest to newest; lines show PC1 and PC2 spectra, shaded bands show leave-one-buoy-out IQRs, and vertical dashed lines mark the local Coriolis frequency from campaign mean latitude.

**Table 5** Cumulative variance explained by PC1 and PC2 for each Antarctic campaign over the common overlapping analysis window.

| Campaign | PC1 % variance | PC2 % variance | Cumulative PC1 to PC2 % variance |
|---|---|---|---|
| SIPEX II BUOYS 2012 | 86.7% | 4.9% | 91.6% |
| SIPEX II WAVES 2012 | 56.5% | 25.8% | 82.3% |
| SCALE SPRING 2019 | 60.2% | 27.3% | 87.5% |
| SCALE WINTER 2019 | 69.6% | 16.6% | 86.1% |
| SHARC 2022 | 63.5% | 25.7% | 89.2% |
| STiMPI 2000 | 67.1% | 13.8% | 80.9% |
| WIIOS 2017 | 98.7% | 1.3% | 100.0% |
| PIPERS 2017 | 65.6% | 14.1% | 79.7% |

Here, it can be seen from columns 2–4 that the Antarctic campaigns are more energetic than the Arctic by an order of magnitude: the total resolved variance for the SO is 0.0379 $(\mathrm{m/s})^2$ in comparison to 0.0035 $(\mathrm{m/s})^2$ and 0.0031 $(\mathrm{m/s})^2$ for the BG and TD regimes of the Arctic respectively. In the Antarctic, the CCW variance is approximately double the variance of the CW variance due to the influence of inertial oscillations while in the Arctic, the CW variance is greater than the CCW accordingly. Of note is the fact that the BG and TD regimes capture similar variance in their spectra despite the differing currents driving their low frequency motion. All three comparison sets remain low-frequency dominated, but BG is the most low-frequency concentrated with approximately 87% of variance captured below 0.8 cpd, TD is the least (approximately 79%), and the Antarctic campaign median is intermediate but still strongly weighted to low frequency at approximately 70%. Near $f_c$, the TD regime and the Antarctic campaign set both retain more total variance than the BG regime (8.8% and 16.4% versus 3.0%, respectively), but the Antarctic campaign median does not retain as much high-frequency variance above 2 cpd as TD (4.5% versus 6.7%). It is clear that all three comparison sets are strongly rotary-asymmetric in the Coriolis frequency band, consistent with a substantial inertial contribution while recognising that tidal and internal-ice-stress variability may overlap this band. Finally, the sub-inertial frequency amplitude drop off across the three regions is broadly similar, ranging from −1.62 to −1.76, with BG showing the steepest decay and TD the shallowest.

### 6.1.1 Area-normalised spectral shape

Spectral shape similarity of the intertial-sense dominant spretra are now compared between the regions after area normalisation. The pairwise shape distances, from Section 3.2.1.1, are shown in Fig. 8.

First the BG and TD regimes are compared with a normalised distance of 0.23. In the Antarctic, the spectral shape of the campaigns show spectral shape distances between 0.16 (e.g. between STiMPI 2000 and SIPEX II Buoys 2012) and 0.74 (e.g. between SCALE Spring 2019 and SHARC 2022). I.e.

**Table 6** Inter-regional rotary comparison summary for the two Arctic regimes and the Antarctic campaign set. Integrated variance is evaluated over 0.1–6 cpd; the band fractions denote the fraction of total resolved rotary variance below 0.8 cpd, within a ±0.25 cpd window around the local $f_c$, and above 2 cpd. The near-$f_c$ dominant-branch fraction denotes the fraction of near-$f_c$ variance carried by the inertial-sense branch (CW in the Arctic, CCW in the Antarctic).

| | BG | TD | Antarctic campaigns |
|---|---|---|---|
| Cases | 33 | 33 | 8 |
| Median CW variance $\left(\frac{m}{s}\right)^2$ | 0.0018 | 0.0017 | 0.0114 |
| Median CCW variance $\left(\frac{m}{s}\right)^2$ | 0.0013 | 0.0011 | 0.0231 |
| Median total variance $\left(\frac{m}{s}\right)^2$ | 0.0035 | 0.0031 | 0.0379 |
| Low-band variance (%) | 86.8 | 79.5 | 69.9 |
| Near-$f_c$ variance (%) | 3.0 | 8.8 | 16.4 |
| High-band variance (%) | 2.7 | 6.7 | 4.5 |
| Near-$\lvert f \rvert$ dominant branch (%) | 92.6 | 90.8 | 96.3 |
| Median sub-inertial slope (–) | −1.76 | −1.62 | −1.67 |

in the Antarctic the STiMPI 2000 campaign exhibits far more similarity of variance distribution with SIPEX II Buoys 2012 than between SCALE Spring 2019 and SHARC 2022. Several Antarctic campaigns are actually closer in variance distribution to the Arctic than the Arctic regions are to themselves: STiMPI 2000 and SHARC 2022 when compared to the TD regime show shape distances of 0.20 and 0.22 respectively.

### 6.1.2 Absolute spectral amplitude differences

Following from this the RMSD between the SO campaigns and Arctic regimes are compared in Fig. 9 below. Where before, Fig. 8 compared the similarity in how variance in the spectra was distributed between the campaigns, now the comparison reflects the magnitudes of energy involved with that distribution.

Here, the Arctic regimes show a small difference in amplitudes of 1 $(\mathrm{cm/s})^2/\mathrm{cpd}$. Within the SO, amplitude comparisons always exceed this with a minimum RMSD of 8 $(\mathrm{cm/s})^2/\mathrm{cpd}$ between STiMPI 2000 and SHARC 2022. On the other side of the scale, SCALE Spring 2019 is the most dissimilar to other SO spectra in terms of amplitude with a maximum RMSD of 168 $(\mathrm{cm/s})^2/\mathrm{cpd}$ between itself and SIPEX II Buoys 2012. SCALE Spring 2019 is also the most different to the Arctic with an RMSD of

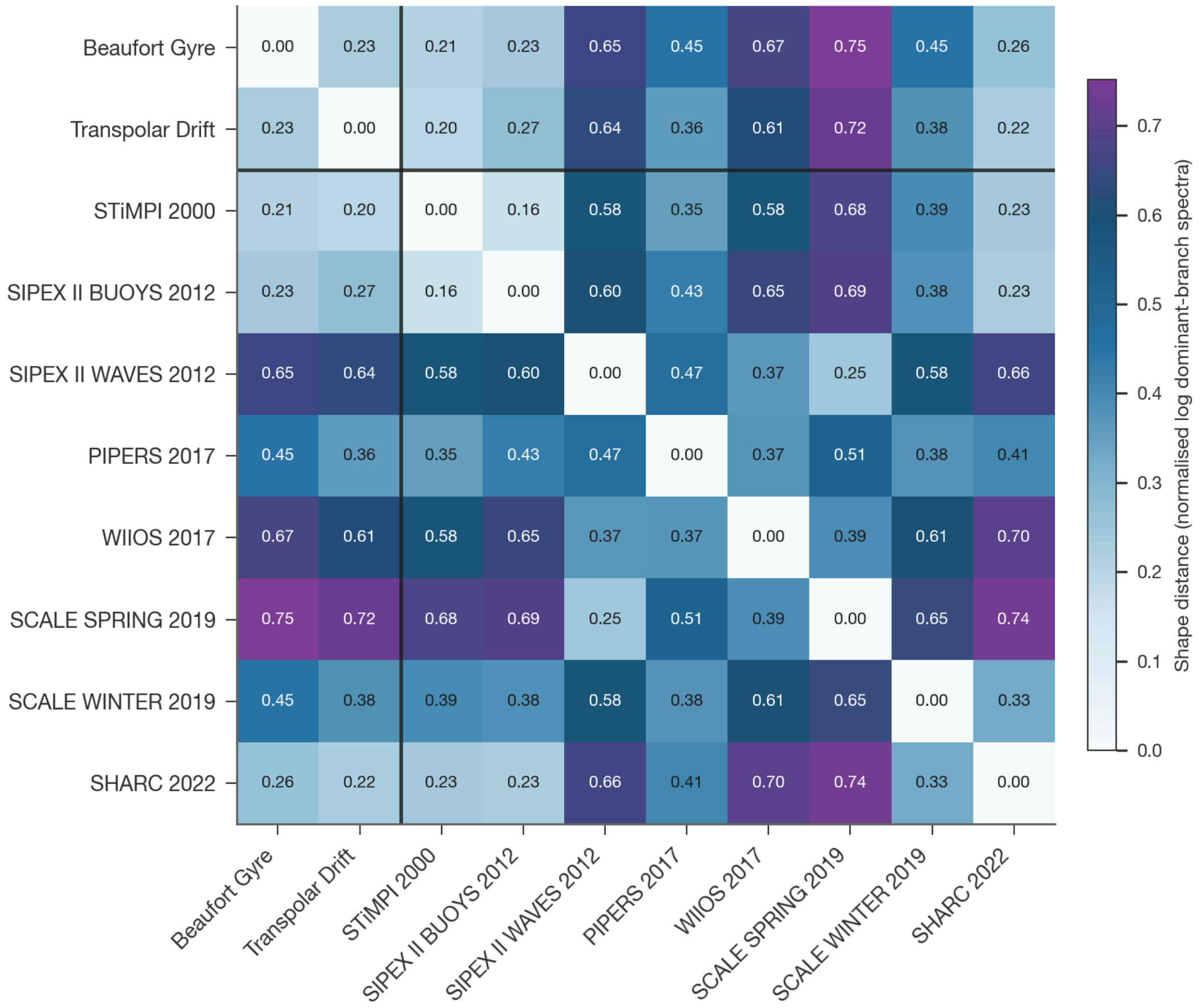

**Fig. 8** Pairwise spectral shape distance across BG, TD, and Antarctic campaigns. Distances are computed from area-normalised log dominant-branch spectra on a common frequency grid (CW in the Arctic, CCW in the Antarctic), so lower values indicate more similar spectral envelopes independent of absolute amplitude.

169 $(\mathrm{cm/s})^2/\mathrm{cpd}$ with both the BG and TD regimes. SIPEX II Buoys 2012 is the most similar to the Arctic at 3 $(\mathrm{cm/s})^2/\mathrm{cpd}$ for both comparisons.

Taken together, Fig. 8 and Fig. 9 indicate that the strongest cross-polar similarity lies in spectral ordering and band structure rather than in absolute spectral amplitude. This is clearest for STiMPI 2000 and SHARC 2022: their normalised shape distances to the Arctic are low and comparable to the Arctic-Arctic comparison itself, with values of 0.20 and 0.22 against TD and 0.21 and 0.26 against BG, versus 0.23 between BG and TD. However, their raw amplitude-sensitive RMSD values remain much larger than the Arctic-Arctic value of 1 $(\mathrm{cm/s})^2/\mathrm{cpd}$, reaching 18 $(\mathrm{cm/s})^2/\mathrm{cpd}$ and 15 $(\mathrm{cm/s})^2/\mathrm{cpd}$ against TD and 18 $(\mathrm{cm/s})^2/\mathrm{cpd}$ and 16 $(\mathrm{cm/s})^2/\mathrm{cpd}$ against BG. Thus, Antarctic campaigns distribute variance across low-frequency, near-inertial, and higher-frequency bands in a way that resembles the Arctic, but they do not generally match the Arctic in absolute spectral level.

SIPEX II Buoys 2012 is the main exception to this amplitude contrast and therefore stands out as an amplitude outlier within the Antarctic set. Its raw RMSD to the Arctic is only 3 $(\mathrm{cm/s})^2/\mathrm{cpd}$ against both BG and TD, much closer to the Arctic-Arctic value of 1 $(\mathrm{cm/s})^2/\mathrm{cpd}$ than to the larger Antarctic comparisons. Yet its normalised shape distances to BG and TD are 0.28 and 0.26 respectively, which are not as low as the corresponding values for STiMPI 2000 and SHARC 2022. In other words, SIPEX II Buoys 2012 is unusual because its absolute spectral level is Arctic-like, whereas STiMPI 2000 and SHARC 2022 are notable because their spectral envelopes are Arctic-like. A plausible explanation is its location closer to the Antarctic continent, where more consolidated or mechanically constrained ice could reduce drift variance relative to more mobile Antarctic campaigns; however, the present analysis does not include an independent ice-concentration or distance-to-coast test. This also supports treating spectral shape similarity and amplitude similarity as distinct aspects of the inter-regional comparison.

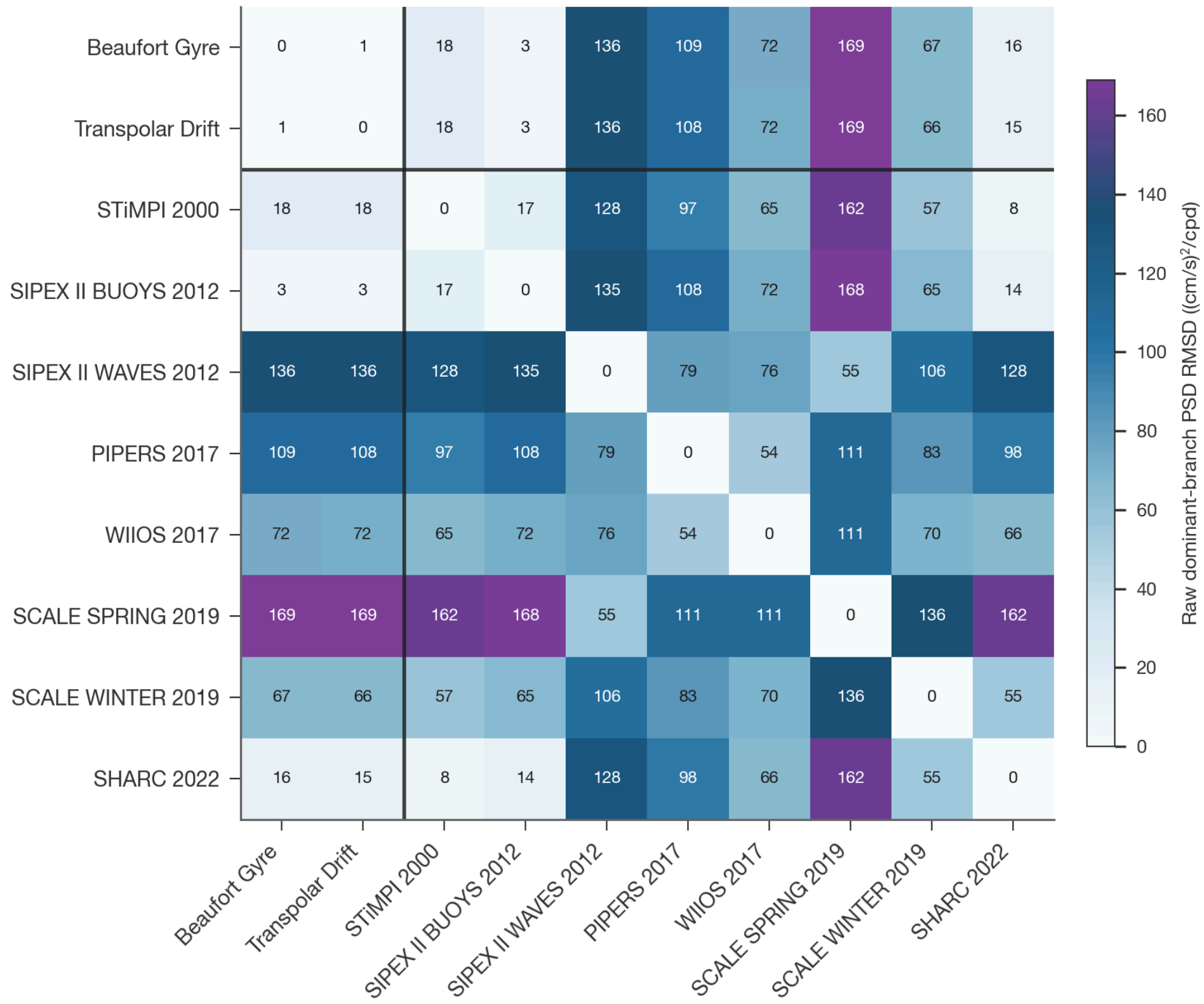

**Fig. 9** Raw amplitude-sensitive RMSD of dominant-branch rotary PSD across BG, TD, and Antarctic campaigns (CW in the Arctic, CCW in the Antarctic). Lower values indicate more similar absolute spectral energy levels over the shared frequency grid.

## 6.2 Inter-buoy coherence across regions

Within-array inter-buoy coherence is compared across the BG, TD, and Antarctic campaign sets in Fig. 10 below. These curves are summaries of pairwise coherence between buoy velocity time series, not coherence between regional median spectra. For each pair, coherence is computed from the overlapping $u$ and $v$ component records and then averaged between components as described in Section 3.2.2.

The regional curves above summarise pairwise estimates by frequency. The ensemble Antarctic coherence curve is only shown where at least half of the campaigns support the frequency bin. This prevents edge artefacts that appear when only the longest campaigns contribute below $\approx 0.3$ cpd. This does not indicate negligible coherence in the Antarctic at those frequencies, rather, it indicates that most Antarctic campaigns are too short to provide a stable pooled estimate there.

Here, it is clear that pairwise coherence is highest at the longest timescales and declines toward higher frequencies in BG, TD, and the SO campaign set. The Arctic regimes show lower coherence in general than in the SO and are constrained to values between 0.2–0.4 across all frequencies with notable rises at the mean inertial frequency for the Arctic. These overall low coherence values are likely reflective of the greater spatial heterogeneity and different effective averaging when compared to the SO campaigns.

The SO ensemble median shows substantially higher pairwise coherence than either Arctic regime. From approximately 0.3–1 cpd, a large proportion of the motion energy at these frequencies is organised coherently across the various buoys with coherence values above 0.6. Coherence decreases with frequency in general with a rise in the local inertial band. These overall higher coherence values are, however, likely not showing that ice motion in the Antarctic is intrinsically more coherent. They most likely reflect the smaller, more compact arrays and shorter common-overlap windows rather than the longer, larger Arctic arrays with greater spatial separation. In the Antarctic each campaign is constrained to sample a smaller area and thus samples more similar local forcing and ice states than in the Arctic, leading to greater coherence. Thus, this cross-region comparison should only be used to show that motion is most coherent at low frequencies and becomes progressively less shared toward the high-frequency band, where local ice mechanics, wave effects, and ocean-coupled variability become more important within the sampled array geometry.

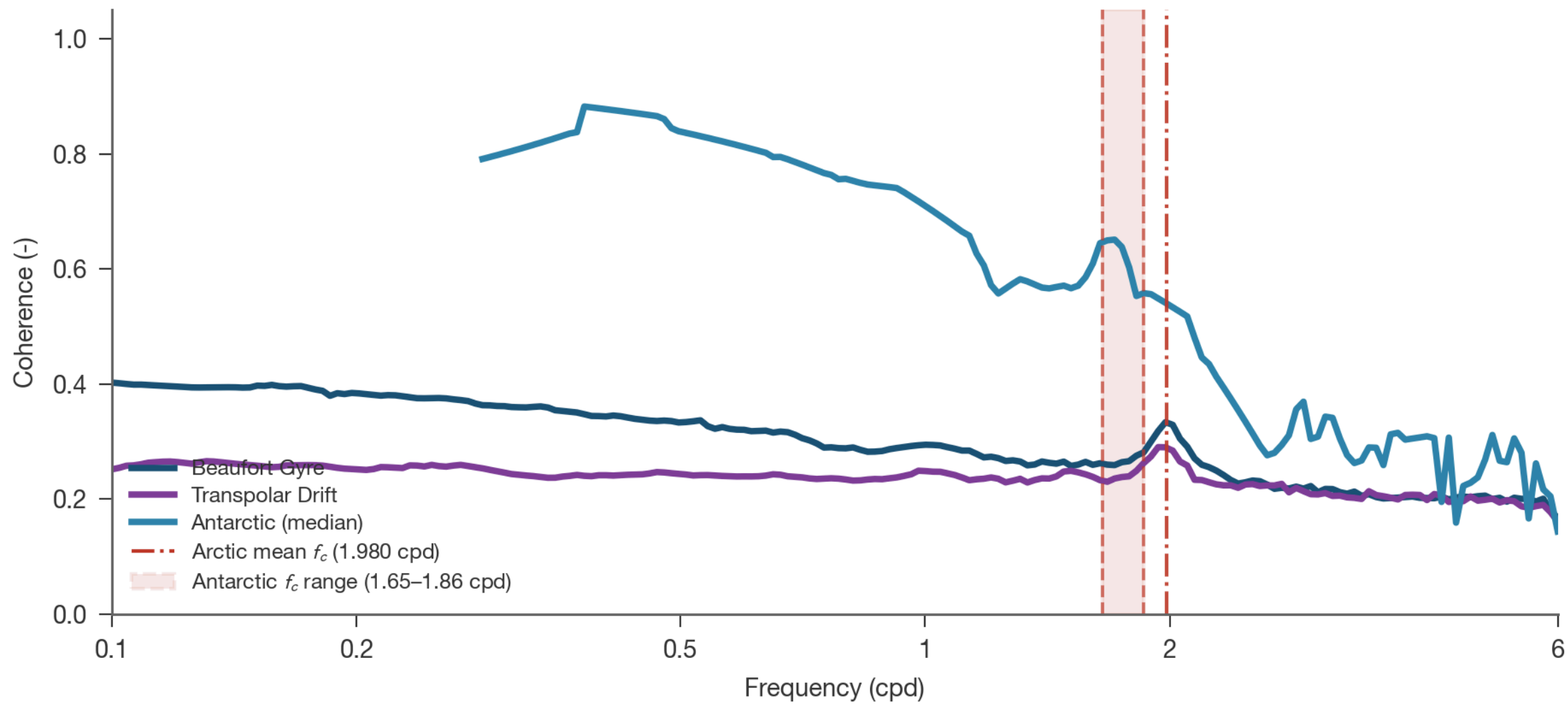

**Fig. 10** Median within-array inter-buoy coherence versus frequency for BG, TD, and the Antarctic campaign set. Lines show regional median coherence; the dash-dotted vertical line marks the Arctic mean $f_c$, and the shaded band marks the Antarctic $f_c$ range.

### 6.3 Cross-regional PCA mode structure

Finally, the inter-region comparison concludes by comparing the spectra of the first two PCs of the region ensembles. Fig. 11 shows the PC1 and PC2 comparisons for the BG, TD, and Antarctic campaign sets.

The left and right of Fig. 11 provide a supportive comparison of modal compactness across the regions rather than a primary inter-regional similarity metric. In the left panel, PC1 is low-frequency dominated in BG, TD, and the Antarctic campaign set, indicating that the leading mode in both oceans represents coherent array-scale drift concentrated at the longest timescales. The BG and Antarctic PC1 median spectra are notably similar at the lowest resolved frequencies, approaching $10^{-1}$ $(\mathrm{m/s})^2/\mathrm{cpd}$ at $\approx 0.1$ cpd, while TD is somewhat larger, approaching $10^{0}$ $(\mathrm{m/s})^2/\mathrm{cpd}$ at its lowest resolved frequencies. In general, PC1 is weakest for the BG regime spectra of the regions. All three spectra then decay strongly with increasing frequency but retain a secondary enhancement near the Coriolis band, with TD showing the strongest enhancement at these frequencies. This indicates that the leading Antarctic mode occupies the same broad spectral band as the Arctic leading modes rather than representing a fundamentally different timescale class.

The right panel of Fig. 11 shows that PC2 retains the same broad low-frequency structure as PC1 but carries relatively more variance toward the Coriolis band than PC1. BG remains the most compact and weakest of the three comparison sets, whereas TD shows the largest near-$f_c$ enhancement, reaching about $1.5 \times 10^{-2}$ $(\mathrm{m/s})^2/\mathrm{cpd}$ near 2 cpd. The Antarctic PC2 median spectrum again falls between the Arctic regimes in overall level and exhibits a clear near-inertial enhancement around 1.8 cpd, while still decaying toward higher frequency in the same broad manner. Taken together, the two panels of Fig. 11 suggest that the first two PCA modes in both oceans occupy similar broad spectral bands: PC1 captures the dominant coherent low-frequency drift mode, while PC2 reflects a secondary mode with relatively greater inertial-band influence.

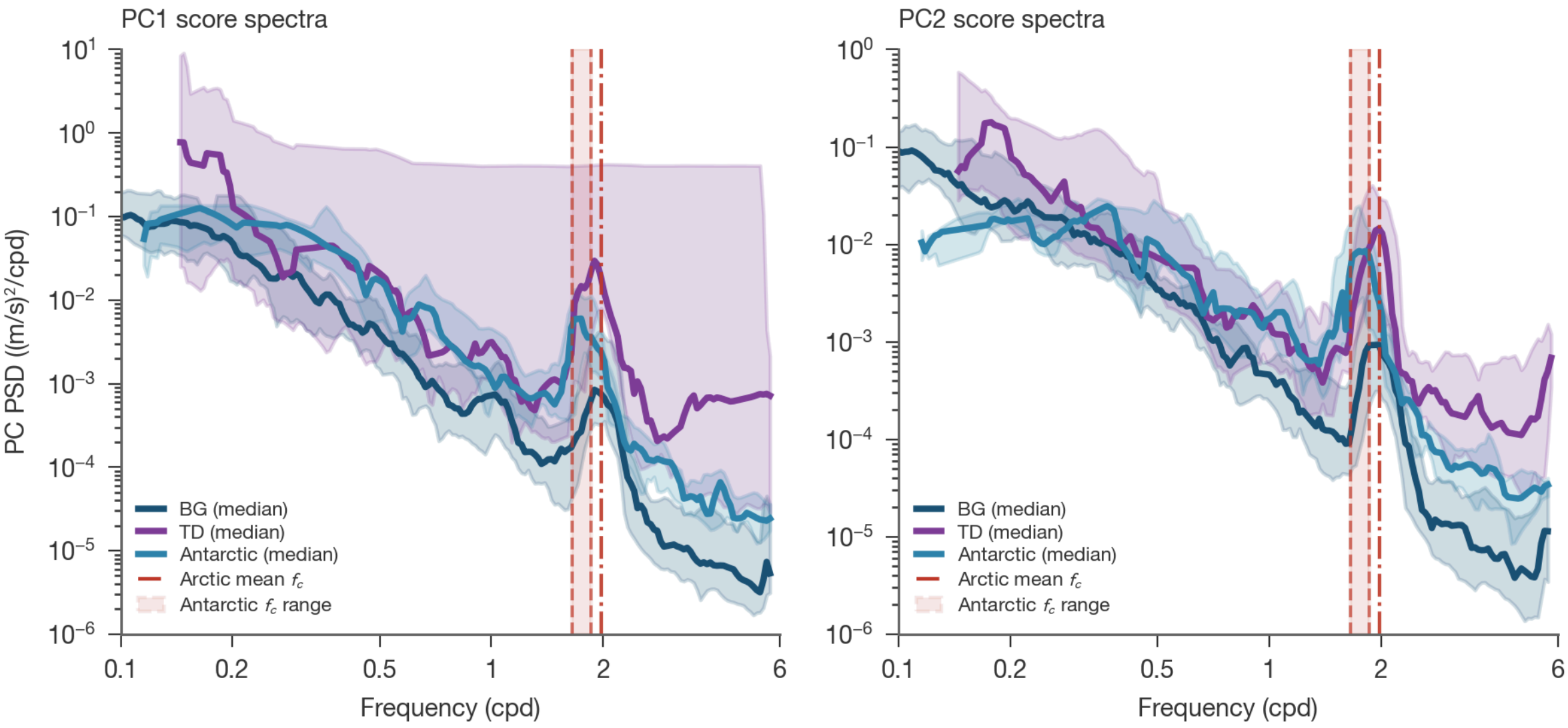

**Fig. 11** PC score-spectrum comparisons across BG, TD, and the Antarctic campaign set. The left panel shows PC1 and the right panel shows PC2. Lines show regional median spectra, shaded bands show IQR or inter-campaign spread, and Coriolis markers show the Arctic mean $f_c$ and Antarctic $f_c$ range.

These PCA comparisons should nevertheless be interpreted cautiously. The modal envelopes remain broad, especially for the first mode of the TD regime, indicating substantial campaign-to-campaign variability, and PCA depends strongly on array geometry, buoy count, and overlap duration. Accordingly, Fig. 11 is used here only to show that the leading within-array modes occupy similar broad spectral bands across regions.

## 7 Discussion

The preceding sections analysed sea ice drift within the Arctic and Antarctic and compared the two regions. Rotary spectral analysis of seasonal drift in the BG and TD regimes showed a shared Arctic timescale structure, with low-frequency dominance and a secondary enhancement near the local inertial-semidiurnal band. In the BG regime, the absence of comparable CCW peaks is consistent with a stronger inertial contribution to the CW branch, while peaks in both the CW and CCW spectra for the TD regime indicate a more mixed inertial-semidiurnal response in which tides, internal ice stresses, and inertial motion may overlap. In the SO, rotary spectra from the various campaigns showed greater variation in amplitudes, spectral ordering and IQR than in the Arctic. In general, however, they also share a similar spectral ordering: low-frequency dominance with decreasing power toward higher frequencies. Absolute energy levels differ between the regions, with Antarctic campaigns often approaching an order of magnitude larger amplitudes than those in the Arctic. Drift spectra also varied in their near-Coriolis enhancement. Some campaigns (e.g. SIPEX II Waves 2012, WIIOS 2017, SCALE Spring 2019) showed large CCW enhancements at these frequencies with no corresponding CW peak, indicating strong inertial susceptibility, while others showed this to a lesser extent; SIPEX II Buoys 2012 showed enhancements to a similar extent in both spectra, indicating a more mixed semidiurnal-band rotary response. Thus, the cross-polar comparison supports similarity in broad spectral organisation and timescale structure, but not equivalence in force balance, internal stress, ocean drag, or wave influence between the two oceans. The high inter-buoy coherence at synoptic and sub-inertial scales is consistent with the classic wind-drift findings of Thorndike and Colony [23], who showed that large-scale ice motion is strongly wind-driven. The present rotary-spectral framework extends those results by quantifying the cross-polar similarity of spectral shape into the inertial-semidiurnal band, where wind, tide, and ice-mechanical influences overlap, and by providing a frequency-resolved benchmark that can be transferred between hemispheres.

The PCA results also support this interpretation, showing that the leading modes in both oceans occupy similar broad spectral bands, with low-frequency dominance and a secondary inertial enhancement. The drift vectors also exhibit relatively low dimensionality over the sampled windows, with the cumulative variance captured by the first three modes in the Arctic accounting for between 55% and 75% of the total variance (from regime-season mean buoy counts of 48-161). Similarly in the SO, the first two modes account for between 80 and 100% of the total variance, but these values are inflated by small buoy counts and short common-overlap windows and are therefore not directly comparable to the larger Arctic ensembles. This interpretation is more meaningful in the Arctic, where larger buoy numbers, broader spatial coverage, and longer time scales strengthen the inference. For experiments whose primary objective is synoptic-scale drift monitoring, this sampled-window low dimensionality suggests that modest buoy counts or wider spacing may be plausible design options to test within mission-specific OSSEs or array-design tradeoff studies where drift field reconstruction accuracy must be set against engineering costs for fixed budget campaigns. More broadly, the required buoy count remains objective-dependent, with denser arrays still needed for inertial variance, deformation, waves, ocean coupling, ice-edge structure, and other smaller-scale heterogeneity. Indeed, the various data sources in this work come from experiments covering a variety of scientific objectives and so the PCA results should be interpreted at a meta-level only for future experiments targeting drift investigations. This is especially important in the Antarctic where data is sparse and so no single buoy-spacing or array-optimisation rule should be over-generalised across mission types despite the low-dimensional array-scale drift representation apparent in the current data.

PCA results should also be interpreted cautiously as supporting metrics, given several limitations. In the Arctic, the IQR of PC1 in the TD regime is very large as evident in the lower row of Fig. 4.

This suggests that the TD regime has high-dimensional variability within its regime-season breakdown and that future work should investigate this more closely for drift-capturing experiments planned in the TD regime. For the SO, datasets are sparse, limited in extent, and short in duration; the high PC1+PC2 variance should be read as low dimensionality over the sampled campaign windows rather than as evidence that larger or longer Antarctic arrays would necessarily show the same dimensionality.

The subsequent inter-regional comparison showed the strongest Arctic-Antarctic similarity after area normalisation, in which Fig. 8 shows that some Antarctic campaigns are as close to BG or TD in terms of the spectral shape as BG and TD are to each other when the hemisphere-appropriate inertial-branch spectra are compared (e.g. STiMPI 2000 and SHARC 2022 when compared to BG and TD). This indicates that some SO campaigns share Arctic-like spectral envelopes and variance ordering across frequency despite the geographic and climatic differences between the regions. It does not show that those campaigns are driven in the same way as the Arctic. Raw RMSD results in Fig. 9 then separate the shape from the amplitudes of the spectra and show that while Antarctic campaigns can look Arctic-like in spectral envelope they remain far more energetic in absolute variance. In terms of energy content in the spectra, overall, only the SIPEX II Buoys campaign in 2012 reflected a similar energy level to the BG and TD regimes (RMSD = 3 $(\text{cm/s})^2$/cpd), whilst the SCALE Spring 2019 campaign was vastly more energetic with an RMSD of 169 $(\text{cm/s})^2$/cpd.

## 8 Conclusion

Given the need for increased in-situ measurements to understand the rapidly changing Arctic environment [44], [45], optimising the deployment of buoys to obtain high-quality data across a wide area is crucial. This is also particularly important in the SO, which significantly impacts Earth's major thermal, climate, and environmental systems, but is one of the least observed environments due to its extreme remoteness and harsh conditions [11], [46]. The analyses presented here aimed to use the drift data from a large range of ice-tethered buoys in the Arctic and SO to identify properties of sea ice drift that can be used to inform new drift experiment design and their GNSS sampling.

These analyses provide a set of frequency-resolved benchmarks, i.e. dominant variance below 0.5 cpd and in the near-inertial band, an expanded Arctic BG/TD seasonal reference, and an explicit separation of spectral-shape similarity from absolute energy-amplitude similarity, that can inform sensor and experiment design. By keeping Antarctic campaigns distinct from the Arctic reference, the comparison offers experiment planners a way to judge whether a proposed observing strategy is likely to capture the same spectral bands as established Arctic deployments, without assuming that the two regions share identical energy levels or dynamical regimes.

The results presented in this paper indicate similarities between the Arctic and SO from the perspective of the frequency domain and dimensionality of sea ice drift. It can be said that the strongest Arctic-SO similarity is in spectral structure, not in spectral amplitude. In both regions, variance is concentrated at low frequencies below 0.5 cpd and in the near-inertial to semidiurnal band where inertial motion, tides, and internal ice stresses may overlap. Thus, from a temporal standpoint, sampling of GNSS location should be done at a frequency of at least 4 cpd in order to satisfy the Nyquist criteria for sampling the two dominant features in the frequency domain. SO campaigns in regions where ice is more mobile, wave-influenced, and less consolidated, are generally higher-energy and more variable than the Arctic regime composites.

The coherent array-scale drift sampled in these regions often exhibits low dimensionality, with a dominant first component, and shows positive coherence between buoy deployment locations. Coherence was shown to be stronger in the SO than the Arctic and is likely due to reduced survivability in the harsher environment, and consequent shorter campaign lengths, along with the smaller spatial regions covered by buoy arrays. Given the need to optimise campaign parameters such as number of buoys, spatial separation, energy use and engineering budgets, the observed coherence and sampled-window dimensionality should be used as inputs to mission-specific OSSEs and array-design tradeoffs for experiments targeting synoptic-scale drift motion. In such studies, wider spacing or reduced buoy counts may be candidate options to evaluate against reconstruction accuracy, survivability, and cost. These options should not be generalised to experiments targeting deformation, waves, ocean coupling, ice-edge structure, inertial-band variability, or smaller-scale heterogeneity, for which dense arrays remain necessary.

A limitation of the analysis in this paper is that the analytical methods used are linear and so cannot capture non-linear relationships in and between the regions in both the temporal and spatial domains, future work will explore nonlinear methods such as kPCA, ICA and autoencoders. Further, because the SO data are limited, it was not possible to investigate whether the findings vary over seasonal timescales; there simply are not enough data to do so. Accordingly, as time progresses and additional data is collected, it is planned to revisit this analysis to investigate whether the findings remain consistent.

## References

[1] B. Lund *et al.*, ‘Arctic Sea Ice Drift Measured by Shipboard Marine Radar’, *J. Geophys. Res. C: Oceans*, vol. 123, no. 6, pp. 4298–4321, Jun. 2018, doi: 10.1029/2018jc013769.

[2] A. S. Thorndike, ‘Diffusion of Sea Ice’, *J. Geophys. Res.*, vol. 91, no. C6, p. 7691, 1986, doi: 10.1029/jc091ic06p07691.

[3] P. Rampal, J. Weiss, D. Marsan, and M. Bourgoin, ‘Arctic Sea Ice Velocity Field: General Circulation and Turbulent-like Fluctuations’, *J. Geophys. Res.*, vol. 114, no. C10, Oct. 2009, doi: 10.1029/2008jc005227.

[4] N. F. Tandon, P. J. Kushner, D. Docquier, J. J. Wettstein, and C. Li, ‘Reassessing Sea Ice Drift and Its Relationship to Long-term Arctic Sea Ice Loss in Coupled Climate Models’, *J. Geophys. Res. C: Oceans*, vol. 123, no. 6, pp. 4338–4359, Jun. 2018, doi: 10.1029/2017jc013697.

[5] P. Heil *et al.*, ‘Tidal Forcing on Sea-Ice Drift and Deformation in the Western Weddell Sea in Early Austral Summer, 2004’, *Deep Sea Res. Part 2 Top. Stud. Oceanogr.*, vol. 55, no. 8–9, pp. 943–962, Apr. 2008, doi: 10.1016/J.DSR2.2007.12.026.

[6] P. Rampal *et al.*, ‘On the Multi-Fractal Scaling Properties of Sea Ice Deformation’, *cryosphere*, vol. 13, no. 9, pp. 2457–2474, Sep. 2019, doi: 10.5194/tc-13-2457-2019.

[7] J. Weiss, D. Marsan, and P. Rampal, ‘Space and Time Scaling Laws Induced by the Multiscale Fracturing of The Arctic Sea Ice Cover’, presented at the IUTAM Symposium on Scaling in Solid Mechanics, Springer Netherlands, 2009, pp. 101–109. doi: 10.1007/978-1-4020-9033-2_10.

[8] W. J. Emery, A. C. Thomas, M. J. Collins, W. R. Crawford, and D. L. Mackas, ‘An Objective Method for Computing Advective Surface Velocities from Sequential Infrared Satellite Images’, *J. Geophys. Res.*, vol. 91, no. C11, p. 12865, 1986, doi: 10.1029/jc091ic11p12865.

[9] H. Sumata *et al.*, ‘An Intercomparison of Arctic Ice Drift Products to Deduce Uncertainty Estimates’, *J. Geophys. Res. C: Oceans*, vol. 119, no. 8, pp. 4887–4921, Aug. 2014, doi: 10.1002/2013jc009724.

[10] S. Bouillon and P. Rampal, ‘On Producing Sea Ice Deformation Data Sets from SAR-derived Sea Ice Motion’, *The Cryosphere*, vol. 9, no. 2, pp. 663–673, 2015, doi: 10.5194/tc-9-663-2015.

[11] L. Newman *et al.*, ‘Delivering Sustained, Coordinated and Integrated Observations of the Southern Ocean for Global Impact’, *Frontiers in Marine Science*, vol. 6, no. JUL, p. 433, Aug. 2019, doi: 10.3389/fmars.2019.00433.

[12] S. Swart *et al.*, ‘Constraining Southern Ocean Air-Sea-Ice Fluxes Through Enhanced Observations’, *Frontiers in Marine Science*, vol. 6, no. JUL, p. 421, Jul. 2019, doi: 10.3389/fmars.2019.00421.

[13] M. H. Derkani *et al.*, ‘Wind, Waves, and Surface Currents in the Southern Ocean: Observations from the Antarctic Circumnavigation Expedition’, *Earth System Science Data*, vol. 13, no. 3, pp. 1189–1209, Mar. 2021, doi: 10.5194/essd-13-1189-2021.

[14] M. Noyce, R. Verrinder, and M. Vichi, 'Identification of Ice Floe Impacts in an Inertial Time Series and Implications for Wave Measurement In Pancake Ice', in *IGARSS 2023 - 2023 IEEE International Geoscience and Remote Sensing Symposium*, Jul. 2023, pp. 149–152. doi: 10.1109/IGARSS52108.2023.10281906.

[15] G. C. Smith *et al.*, 'Polar Ocean Observations: A Critical Gap in the Observing System and Its Effect on Environmental Predictions from Hours to a Season', *Frontiers in Marine Science*, vol. 6, no. JUL, p. 429, Aug. 2019, doi: 10.3389/fmars.2019.00429.

[16] S. R. Smith, P. J. Hughes, and M. A. Bourassa, 'A Comparison of Nine Monthly Air-Sea Flux Products', *Int. J. Climatol.*, vol. 31, no. 7, pp. 1002–1027, Jun. 2011, doi: 10.1002/joc.2225.

[17] L. Yu *et al.*, 'An Inter-Comparison of Six Latent and Sensible Heat Flux Products over the Southern Ocean', *Polar Res.*, vol. 30, no. SUPPL.1, Nov. 2011, doi: 10.3402/polar.v30i0.10167.

[18] Y. Fujii *et al.*, 'Observing System Evaluation Based on Ocean Data Assimilation and Prediction Systems: On-Going Challenges and a Future Vision for Designing and Supporting Ocean Observational Networks', *Frontiers in Marine Science*, vol. 6, 2019, doi: 10.3389/fmars.2019.00417.

[19] X. Zeng *et al.*, 'Use of Observing System Simulation Experiments in the United States', *Bull. Am. Meteorol. Soc.*, vol. 101, no. 8, pp. E1427–E1438, 2020, doi: 10.1175/BAMS-D-19-0155.1.

[20] C. A. Geiger, S. F. Ackley, and W. D. Hibler III, 'Sea Ice Drift and Deformation Processes in the Western Weddell Sea', in *Antarctic Sea Ice: Physical Processes, Interactions and Variability*, American Geophysical Union (AGU), 1998, pp. 141–160. doi: 10.1029/AR074p0141.

[21] M. J. Doble and P. Wadhams, 'Dynamical Contrasts between Pancake and Pack Ice, Investigated with a Drifting Buoy Array', *Journal of Geophysical Research: Oceans*, vol. 111, no. C11, Oct. 2006, doi: 10.1029/2005JC003320.

[22] A. Womack, A. Alberello, M. de Vos, A. Toffoli, R. Verrinder, and M. Vichi, 'A Contrast in Sea Ice Drift and Deformation between Winter and Spring of 2019 in the Antarctic Marginal Ice Zone', *The Cryosphere*, vol. 18, no. 1, pp. 205–229, Jan. 2024, doi: 10.5194/tc-18-205-2024.

[23] A. S. Thorndike and R. Colony, 'Sea Ice Motion in Response to Geostrophic Winds', *Journal of Geophysical Research: Oceans*, vol. 87, no. C8, pp. 5845–5852, 1982, doi: 10.1029/JC087iC08p05845.

[24] A. Alberello *et al.*, 'Drift of Pancake Ice Floes in the Winter Antarctic Marginal Ice Zone During Polar Cyclones', *Journal of Geophysical Research: Oceans*, vol. 125, no. 3, p. e2019JC015418, 2020, doi: 10.1029/2019JC015418.

[25] A. Womack, M. Vichi, A. Alberello, and A. Toffoli, 'Atmospheric Drivers of a Winter-to-Spring Lagrangian Sea-Ice Drift in the Eastern Antarctic Marginal Ice Zone', *J. Glaciol.*, vol. 68, no. 271, pp. 999–1013, 2022, doi: 10.1017/JOG.2022.14.

[26] A. L. Kohout, B. Penrose, S. Penrose, and M. J. M. Williams, 'A Device for Measuring Wave-Induced Motion of Ice Floes in the Antarctic Marginal Ice Zone', *Ann. Glaciol.*, vol. 56, no. 69, pp. 415–424, 2015, doi: 10.3189/2015AoG69A600.

[27] M. Di *et al.*, 'GNSS Real–Time Precise Point Positioning in Arctic Northeast Passage', *Journal of Marine Science and Engineering*, vol. 10, no. 10, p. 1345, Oct. 2022, doi: 10.3390/jmse10101345.

[28] J. Toole, R. Krishfield, M.-L. Timmermans, and A. Proshutinsky, 'The Ice-Tethered Profiler: Argo of the Arctic', *Oceanography*, vol. 24, no. 3, pp. 126–135, Sep. 2011, doi: 10.5670/oceanog.2011.64.

[29] R. Krishfield, J. Toole, A. Proshutinsky, and M.-L. Timmermans, 'Automated Ice-Tethered Profilers for Seawater Observations under Pack Ice in All Seasons', *J. Atmos. Ocean. Technol.*, vol. 25, no. 11, pp. 2091–2105, Nov. 2008, doi: 10.1175/2008JTECHO587.1.

[30] M. de Vos, C.-L. Ramjukadh, M. de Villiers, C. Lyttle, A. Womack, and M. Vichi, 'Polar Iridium Surface Velocity Profilers (p-iSVP), and Standard Iridium Surface Velocity Profilers (iSVP) during SCALE 2019 Winter and Spring Cruises [Data Set]'. 2023.

[31] A. L. Kohout, M. Smith, L. A. Roach, G. Williams, F. Montiel, and M. J. M. Williams, 'Observations of Exponential Wave Attenuation in Antarctic Sea Ice during the PIPERS Campaign', *Ann. Glaciol.*, vol. 61, no. 82, pp. 196–209, Sep. 2020, doi: 10.1017/aog.2020.36.

[32] A. Kohout, M. Williams, M. Smith, G. Williams, L. Roach, and S. Ackley, *PIPERS Waves in Ice Observations*. (2021). Mendeley Data. doi: 10.17632/b33xxnpfyw.1.

[33] K. Machutchon *et al.*, 'SA Agulhas II Winter 2017 Cruise: Ice Edge Drifters'. 2019.

[34] C. Eayrs *et al.*, *SA Agulhas II Winter 2017 Cruise: Waves In Ice Observation Systems (WIIOS), Ver. 1.* (2019). Australian Antarctic Data Centre. doi: 10.26179/5cc934992f065.

[35] P. Heil, J. Hutchings, R. Stevens, and O. Lecomte, 'In Situ Lagrangian Drifting Buoy Data off East Antarctica for the Austral Spring of 2012, Deployed during the SIPEX II Voyage of the Aurora Australis'. 2020.

[36] International Arctic Buoy Programme, 'Data Products'. Accessed: Sep. 06, 2024. [Online]. Available: https://iabp.apl.uw.edu/Data_Products/

[37] A. Bliss *et al.*, 'Sea Ice Drift Tracks from the Distributed Network of Autonomous Buoys Deployed during the Multidisciplinary Drifting Observatory for the Study of Arctic Climate (MOSAiC) Expedition 2019 - 2021', 2022, doi: 10.18739/A2KP7TS83.

[38] M. Hoppmann *et al.*, *Raw Data of GPS Position and Sea-Surface Temperature Recorded by 44 Southtek NOMAD Drifting Buoys in the Marginal Ice Zone North of Svalbard in 2022*. (2023). PANGAEA. doi: 10.1594/PANGAEA.959001.

[39] P. Elosegui *et al.*, 'Sea Ice Drift Tracks for the Sea Ice Dynamic Experiment (SIDEx) Field Campaign from Geodetic Global Navigation Satellite Systems (GNSS) SATICE Buoys, Alaska, 2021', 2023, doi: 10.18739/A2DF6K50X.

[40] J. Hutchings *et al.*, 'Sea Ice Drift Tracks for The Sea Ice Dynamic Experiment (SIDEx) Field Campaign from Global Navigation Satellite Systems (GNSS) Ice Trackers, Alaska, 2021', 2023, doi: 10.18739/A2J678Z4N.

[41] J. Gonella, 'A Rotary-Component Method for Analysing Meteorological and Oceanographic Vector Time Series', *Deep Sea Research and Oceanographic Abstracts*, vol. 19, no. 12, pp. 833–846, Dec. 1972, doi: 10.1016/0011-7471(72)90002-2.

[42] R. E. Thomson and W. J. Emery, 'Chapter 5 - Time Series Analysis Methods', in *Data Analysis Methods in Physical Oceanography (Fourth Edition)*, Fourth Edition., R. E. Thomson and W. J. Emery, Eds, Elsevier Science, 2024, pp. 515–669. doi: 10.1016/B978-0-323-91723-0.00011-8.

[43] M. Athanase, R. Köhler, C. Heuzé, X. Lévine, and R. Williams, 'The Arctic Beaufort Gyre in CMIP6 Models: Present and Future', doi: 10.1029/2024JC021873.

[44] M.-L. Timmermans and J. Marshall, 'Understanding Arctic Ocean Circulation: A Review of Ocean Dynamics in a Changing Climate', *J. Geophys. Res. C: Oceans*, vol. 125, no. 4, Apr. 2020, doi: 10.1029/2018jc014378.

[45] D. Rudnick, D. Costa, K. Johnson, C. Lee, and M.-L. Timmermans, 'ALPS II – Autonomous Lagrangian Platforms and Sensors', 2018.

[46] F. Barbariol *et al.*, 'Large Waves and Drifting Buoys in the Southern Ocean', *Ocean Eng.*, 2019, doi: 10.1016/j.oceaneng.2018.12.011.